\documentclass[numberedappendix]{openjournal}
\usepackage{amsmath}
\usepackage{multirow}
\usepackage{float}

\usepackage{xcolor}
\usepackage{textgreek}
\usepackage[utf8]{inputenc}
\usepackage[english]{babel}

\usepackage{hyperref}
\hypersetup{
    unicode, 
    colorlinks=true,
    linkcolor=linkcolor,
    citecolor=linkcolor,
    filecolor=linkcolor,
    urlcolor=linkcolor,
}
\usepackage{color,colortbl}
\definecolor{linkcolor}{rgb}{0.0,0.3,0.5}
\usepackage{tensind}
\tensordelimiter{?}
\DeclareGraphicsExtensions{.bmp,.png,.jpg,.pdf}
\usepackage{verbatim}
\usepackage[normalem]{ulem}
\usepackage{orcidlink}
\usepackage{soul}
\graphicspath{ {./figs/} }

\begin{document}
\title{The velocity coherence scale: a novel probe of cosmic homogeneity and a potential standard ruler}

\author{Leonardo Giani \orcidlink{0000-0001-6778-1030}}

\affiliation{Swinburne University of Technology, Hawthorn VIC 3122, Australia}
\affiliation{School of Mathematics and Physics, The University of Queensland,
 Brisbane, QLD 4072, Australia}
 \affiliation{OzGrav: The ARC Centre of Excellence for Gravitational Wave Discovery }
 \email{uqlgiani@uq.edu.au}

\author{Cullan Howlett\orcidlink{0000-0002-1081-9410}}
% \email{author2@you.com}
\affiliation{School of Mathematics and Physics, The University of Queensland,
 Brisbane, QLD 4072, Australia}
 \affiliation{OzGrav: The ARC Centre of Excellence for Gravitational Wave Discovery }
\author{Chris Blake \orcidlink{0000-0002-5423-5919}}
% \email{author2@you.com}
\affiliation{Swinburne University of Technology, Hawthorn VIC 3122, Australia}
\affiliation{OzGrav: The ARC Centre of Excellence for Gravitational Wave Discovery }
\author{Ryan Turner \orcidlink{0000-0002-7638-2880}}
% \email{author2@you.com}
\affiliation{The Australian National University, Camberra ACT 2601, Australia}
\affiliation{OzGrav: The ARC Centre of Excellence for Gravitational Wave Discovery }

\author{Tamara M.~Davis \orcidlink{0000-0002-4213-8783}}
% \email{author2@you.com}
\affiliation{School of Mathematics and Physics, The University of Queensland,
 Brisbane, QLD 4072, Australia}
 \affiliation{OzGrav: The ARC Centre of Excellence for Gravitational Wave Discovery }

\begin{abstract}
We introduce the \textit{velocity coherence scale} $R_v$, the scale at which the spherical volume average of the trace of the velocity correlation tensor transitions from scaling faster than the sphere radius to scaling more slowly. This corresponds to the radius at which the average motion of galaxies along their separation vectors transitions from correlated to anti-correlated. More intuitively, $R_v$ represents the scale at which galaxies, on average, cease to move coherently.
We derive a theoretical estimator for $R_v$ by defining the \textit{bulk in spheres} $\mathcal{B}_R$, a velocity-field analogue of the mean scale counts used in density-field correlation analyses. We show that, for a statistically homogeneous matter distribution, the logarithmic derivative of $\mathcal{B}_R$ and the correlation dimension $D_2$ share the same asymptotic behaviour and therefore can be used to estimate the scale of transition to statistical homogeneity. Furthermore, we show that in standard $\Lambda$CDM cosmologies the velocity coherence scale is tightly connected to the matter–radiation equality scale, and that its value in comoving coordinates is redshift-independent. These results highlight the potential of $R_v$ both as a standard ruler and as a physically motivated scale characterising the onset of cosmic homogeneity.

We present a proof of concept using measurements of the peculiar velocity correlation functions from the Sloan Digital Sky Survey. We show that the main challenge in determining $R_v$ is the limited precision of peculiar velocity measurements compared to density ones, as they typically rely on smaller samples with larger uncertainties that scale roughly linearly with survey depth. Fitting our theoretical estimators for $R_v$ with both a parabolic model and a third-order polynomial, we obtain $R_v \approx 138^{+23}_{-54}\,\mathrm{Mpc}/h$. Finally, we show that more precise determinations should be achievable with current and upcoming peculiar velocity surveys.
\end{abstract}

\maketitle

\section{Introduction}\label{Intro}
The cosmological principle posits that the Universe, on sufficiently large scales, is statistically homogeneous and isotropic. Observations broadly support this foundational assumption of the standard cosmological model, yet determining the scales at which deviations from homogeneity may bias cosmological inference remains a subtle and nuanced task.

A number of consistency tests of cosmic homogeneity have been conducted using the three-dimensional distribution of galaxies. One such test is based on the consideration that, in a homogeneous distribution, the number of galaxies contained within a sphere of radius $r$ scales proportionally with the sphere's volume. The radius $R_H$ at which this proportionality is reached (within a specified tolerance) provides an empirical characterisation of the transition to homogeneity \citep{1994ApJ...437..550M,Martinez:1998yp,Amendola:1999gd,Pan:2000yg,Yadav:2005vv,2009MNRAS.399L.128S,2009A&A...496....7S,2009A&A...505..981S,Labini:2010qx,Labini:2011dv,Ntelis:2017nrj}. Different estimators have been considered in the literature \citep{Borgani:1994uy,Martinez:2002mi,Shao:2025xgi,Bizarria:2025jdh}, and it has been shown that the homogeneity scale itself, if measured at different redshifts, can be used as a standard ruler within a given cosmological model \citep{Ntelis:2018ctq,Avila:2021mbj}. However, these analyses have also revealed that the transition scale measured from galaxy samples is degenerate with the galaxy bias $b$, which describes the proportionality between the tracer and the underlying dark matter density field. Consequently, inferring $R_H$ for the matter distribution from galaxy surveys requires assumptions about both the cosmological model and the galaxy bias \citep{Scrimgeour:2012wt}.

In this work we advocate a new method to characterise the transition-to-homogeneity scale using peculiar velocity (PV) measurements. At scales where linear cosmological perturbation theory applies, PVs offer two key advantages: they provide an unbiased ($b$ -independent) tracer of the underlying matter distribution, and they are particularly sensitive to the largest-scale modes. Moreover, recent evidence for anomalously large bulk flows in the local Universe \citep{Aluri:2022hzs,Watkins_2023,Whitford:2023oww,Watkins:2025mlc} provides additional motivation to investigate the transition to homogeneity using PVs as a complementary probe. This is particularly timely in view of upcoming surveys such as the Dark Energy Spectroscopic Instrument (DESI) PV survey and the 4MOST Hemispheric Surveys (4HS), as well as the growing interest in whether backreaction effects from inhomogeneities may help alleviate several current cosmological tensions \citep{Clifton:2024mdy,Giani:2023aor,Giani:2024nnv,Giani:2025hhs,Lane:2023ndt,Camarena:2025upt,Galoppo:2025hzy}.

The core idea of our approach is that inhomogeneities in the density field induce characteristic correlations in the peculiar motions of distant tracers. For example, the gravitational field of a central mass generates anti-correlated motions between galaxy pairs positioned on opposite sides of it, and correlated motions for pairs within the same hemisphere. To analyse the statistical properties of these correlations in a homogeneous matter distribution, we define the \textit{bulk in spheres}, i.e. the spherical average of the trace of the velocity correlation tensor. We then define the \textit{velocity coherence scale} $R_v$ as the scale at which these averaged correlations transition from decreasing more slowly than the sphere radius to decreasing more rapidly than it. Physically, this corresponds to the scale at which, on average, the motion of galaxies within the sphere transitions from positive to negative correlation along their separation vectors. To illustrate the potential of this approach, we apply it to measurements of the PV correlation function from Ref.~\cite{Lyall:2024mks} for the Sloan Digital Sky Survey (SDSS) PV catalogue \citep{Howlett:2022len}, and assess the detectability of the velocity coherence scale with current and upcoming datasets.

The structure of the paper is as follows. In Section~\ref{Theory} we introduce peculiar velocity correlation functions, review the correlation dimension as a probe of the homogeneity scale, and {introduce the velocity coherence scale and its connection with a homogeneous distribution}. In Section~\ref{SDSS} we provide a proof of concept by applying the methodology to measurements of the velocity correlation function from SDSS. Finally, Section~\ref{discussion} presents a discussion of our results and our conclusions.

\section{Theory}\label{Theory}

\subsection{Velocity correlation functions}
If the velocity field is linear, homogeneous, isotropic and irrotational, the correlation between the velocity components $\left(i,j\right)$ of two tracers at positions $A$ and $B$ can be written \citep{gorski,Groth1989,Wang2018,Wang2021,Turner:2022mla,Blake:2023kpk,Turner:2024blz}
\begin{eqnarray}
    &&\langle v_i(\vec{r}^A),v_j(\vec{r}^B)\rangle =\Psi_{ij}(r)\;,
\end{eqnarray}
where we have introduced the velocity tensor
\begin{equation}
        \Psi_{ij}= \left[\Psi_{\parallel}(r) - \Psi_{\perp}(r) \right]\hat{r}^{A}_i\hat{r}^B_j + \Psi_{\perp}(r)\delta_{ij}\;,
\end{equation}
where $r$ is the distance between $A$ and $B$, $\vec{r}^A$ and $\vec{r}^B$ their positions, and the $\Psi_{||,\perp}$'s describe the correlation between components of the velocity parallel and perpendicular to their separation vector respectively. In linear perturbation theory, these depend only on the variance of the density field and the growth rate of structure, with the functional dependence in Fourier space reading simply:
\begin{equation}\label{psipar}
    \Psi_{\parallel}(r)= \frac{H^2 a^2 f^2}{2\pi^2} \int_0^{\infty} dk P(k)\left[j_0(kr) -2\frac{j_1(kr)}{kr} \right]\;, \qquad \quad 
    \Psi_{\perp}(r)= \frac{H^2 a^2 f^2}{2\pi^2} \int_0^{\infty} dk P(k)\frac{j_1(kr)}{kr}\;.
\end{equation}
Since in practice one can only measure the projection of the velocity field along the line of sight $u$, it is useful to define the following functions
\begin{equation}
     \langle \Psi_1(r) \rangle =  \frac{\sum_{A,B}^{r_{\rm{bin}}} u_A u_B \cos{\theta_{AB}}}{\sum_{A,B}^{r_{\rm{bin}}} \cos^2\theta_{AB}} \qquad \quad \langle \Psi_2(r) \rangle = \frac{\sum_{A,B}^{r_{\rm{bin}}} u_A u_B \cos{\theta_A}\cos{\theta_B}}{\sum_{A,B}^{r_{\rm{bin}}} \cos{\theta_{AB}}\cos{\theta_A}\cos{\theta_B}}\;,
\end{equation}
with the sums taken over {all the unique ($A<B$) pairs of galaxies in the $r$ separation bin}, and where the $\theta$'s are the angles between the position and separation vectors of the galaxies ($\cos\theta_i= \hat{r}\cdot \hat{r}_i$, $\cos \theta_{AB}= \hat{r}_A\cdot \hat{r}_B$). Ref.~\cite{gorski} derives the following transformations between $\Psi_{1,2}$ and $\Psi_{\parallel,\perp}$
\begin{equation}
    \langle \Psi_1(r) \rangle = \mathcal{A}(r)\Psi_{\parallel}(r) + \left[1-\mathcal{A}(r)\right]\Psi_{\perp}(r)\;, \qquad \quad \langle \Psi_2(r) \rangle = \mathcal{B}(r)\Psi_{\parallel}(r) + \left[1-\mathcal{A}(r)\right]\Psi_{\perp}(r)\;.
\end{equation}
The functions $A$ and $B$ depend on the survey geometry  
\begin{equation}
    \mathcal{A} = \frac{\sum_{A,B}^{r_{\rm{bin}}} w_A w_B\cos\theta_{AB}}{\sum_{A,B}^{r_{\rm{bin}}} w_A w_B \cos^2\theta_{AB}}\; \qquad \quad \mathcal{B} = \frac{\sum_{A,B}^{r_{\rm{bin}}} w_A w_B\cos\theta_{A}\cos\theta_{B}}{\sum_{A,B}^{r_{\rm{bin}}} w_A w_B \cos\theta_{AB}\cos\theta_{A}\cos\theta_{B}}\;,
\end{equation}
where the weights $w$ account for the distance-dependent error on individual velocity measurements ($w\propto (n_gP_v + \sigma^2)^{-1} $), see for example Refs.~\cite{Qin:2019axr,Turner:2021ddy} for their detailed derivation. 

\subsection{Probing statistical homogeneity with the density field}

Let us briefly describe how the homogeneity scale can be defined in a cosmological setting. We address the reader to appendix~\ref{appendixA} for a more formal discussion of this definition following closely Refs.~\citep{Gabrielli:2000sh,Gabrielli:2001xw}.

A distribution produced by a stationary stochastic process is homogeneous and isotropic if it has a unique, position independent non-vanishing average $\langle\rho(\vec{x})\rangle\equiv \rho_0$ and the order $n$ moments of the distribution depend only on the relative distance between the $n$ points.
It is customary to adopt a threshold value $\lambda$ to identify the transition to homogeneity with the scale $R_\rho$ for which:
\begin{equation}\label{homdefcorf}
\left|\frac{1}{R^3}\int_0^R dr\; r^2 \xi(\vec{x}_o + \vec{r})\right|\leq \lambda \;, \forall R\geq R_\rho, \forall \; \vec{x}_o\; \in \mathcal{V},
\end{equation}
where $\xi(r)$ is the usual 2-point reduced correlation function of the density contrast $\delta(\vec{x}) = (\rho(\vec{x}) - \rho_0)/\rho_0$ and $\vec{x}_o$ the position of any galaxy within the survey volume $\mathcal{V}$. 
An important remark is in order. It is the asymptotic behaviour of the volume average of the correlation function $\xi(r)$\; that establishes whether the underlying density distribution is statistically homogeneous, whilst the choice of the threshold $\lambda$ is somewhat arbitrary. 
In the standard cosmological model, with a standard primordial Harrison Zeldovich spectrum of perturbations $P(k)\propto k^{n}$ (with $n\approx 1$), we have on large scales $\xi(r)\propto r^{-(n+3)}$.

\subsubsection*{The mean scaled counts and the correlation dimension}
{Having defined a suitable definition of homogeneity, let us move on to the description of the most common estimators usually adopted in homogeneity scale measurements: the mean scale counts $\mathcal{N}$ and the correlation dimension $D_2$}.
In the following, barred quantities $\bar{x}$ indicate functions whose definitions rely on homogeneity and isotropy, and are obtained from their theoretical estimators in FLRW cosmologies at linear order in perturbation theory. Non-barred quantities are instead operational functions, computed from observational estimators.
A good estimate of the mean number of neighboring tracers in a sphere of radius $R$ centered on any galaxy is \citep{Peebles1980}:
\begin{equation}\label{numb_count}
    \bar{N}(R) = 4\pi\bar{\rho}\int_0^R d\bar{r} \left[1+\xi(\bar{r})\right]\bar{r}^2\;,
\end{equation}
where $\bar{\rho}$ is the mean density of the tracers and $\bar{\xi}(r)$ their real space two point correlation function. The mean scaled counts is then given by its volume average 
\begin{equation}
    \bar{\mathcal{N}}(R) = \frac{3}{4\pi R^3} \bar{N}(R)\;,
\end{equation}
and whose scaling with the radius essentially defines the correlation dimension $D_2$
\begin{equation}\label{corr_dim}
    \bar{D}_2(R) = 3+ \frac{d \ln \bar{\mathcal{N}}(R)}{d \ln R}\;.
\end{equation}
{Ref.~\cite{Scrimgeour:2012wt} advocated for the use of the following operational estimator for  $\mathcal{N}$ in a given galaxy survey
\begin{equation}\label{NScrimg}
    {\mathcal{N}} = \frac{1}{G}\sum_{i=1}^G \frac{N^i(<R)}{\frac{1}{R}\sum_{j=1}^{n_{\text{rand}}} w_j N_{\text{rand}^{i,j}}(<R)}\;,
\end{equation}
where $G$ is the number of observed tracers, and where the averaged number of objects within a random sphere $N_i$ centered in any of them is normalized by the weighted (with weights $w_j\equiv \rho/\rho_{j,\;\text{rand}}$, where $\rho, \rho_j$ are the densities of galaxies in the survey volume and the $j$th random catalog respectively) averaged number count across $n_{\text{rand}}$ random catalogs. These randoms are generated from a homogeneous distribution sharing the same window functions and number densities of the galaxy survey considered, which as noticed in \citep{Labini:2025dnc} might bias the estimator towards homogeneity.}

{In virtue of Eq.~\eqref{homdefcorf}, for a homogeneous distribution we have $N(R)\propto R^3$, implying $\mathcal{N}(R) \rightarrow 1 $ and $D_2 \rightarrow 3$.} {Since these values are} reached only asymptotically, it is customary (and somewhat arbitrary) to define the homogeneity scale as the one where the correlation dimension and the mean scaled count cross the thresholds $\mathcal{N}=1.01$ and $D_2(R_H)=2.97$, corresponding to $1\%$ deviation from a Poisson distribution.

{Another important remark is in order: both definitions of $\mathcal{N}$ and $\bar{\mathcal{N}}$ (and their logarithmic derivatives) rely on some degree of homogeneity. Indeed, both the definition of a correlation function $\bar{\xi}(r)$ in Eq.~\eqref{numb_count} and the direction independent averaging across all the spheres in Eq.~\eqref{NScrimg} presuppose translational invariance and the existence of a well defined mean density.}\footnote{{Notice that even if the underlying density distribution is inherently direction dependent, nothing prevents one from constructing an isotropic correlation function $\xi(r)$ from the average correlation between tracers with separation $r$. In this case, $\xi(r)$ corresponds to the monopole of the distribution, with higher order moments expected to be non-negligible in a truly inhomogeneous distribution.}} {Therefore, whilst the asymptotic behaviour of these estimators can be used to define the scale of transition to homogeneity scale (within a certain threshold), they cannot be used by themselves to test the cosmological principle. An exception to this is the case of fractal Universes, for which the correlation dimension $D_2$ asymptotes to a different constant  which depends on the value of the fractal dimension $d$.}

\subsection{The ``bulk in spheres'' } 
Let us now define an analogy of the ``count in spheres'' for the velocity field. A sensible choice is to compute the average on a sphere of radius $R$ of what \cite{gorski} refers to as ``total velocity''
% which is nothing but the well known velocity dispersion $\sigma_v$ \cite{Howlett:2019bky}), which we will label \textit{bulk in spheres}:
\begin{equation}\label{bulkinsphere}
    \mathcal{B}_R= \frac{3}{R^3}\int_0^R dr\left(\Psi_\parallel(r)+ 2\Psi_\perp(r)  \right) r^2\;,
\end{equation}
which is particularly interesting because of its connection with bulk flow measurements. Indeed, assuming Gaussian density fluctuations, the velocity field is a gaussian random variate with zero mean and variance that can be calculated from the velocity power spectrum $P_{vv}$ \citep{2012ApJ...761..151L,Andersen:2016ywu}. Smoothing the latter over a sphere of radius $R$ with a uniform window function $W(r)$ 
\begin{equation}
    W(r)=\begin{cases}
        \frac{3}{4\pi R^3} \;\mathrm{if} \; r<R\\
        0 \qquad \mathrm{if}\; r>R
    \end{cases}\;,
\end{equation}
one obtains\footnote{In close analogy with the definition of the rms of the density field $\sigma(R)$, if not for the $k^2$ factor in the integrand and the prefactor $(Hfa)^2$.}
\begin{equation}\label{sigmav}
    \sigma_{v}^2(R)= \frac{H^2f^2a^2}{2\pi^2}\int_0^\infty dk P(k) \tilde{W}(k,R)^2\;, 
\end{equation}
where we have introduced the Fourier transform of the window function
\begin{equation}
    \tilde{W}(k,R)= 4\pi\int_0^\infty dr \; r^2 W(r) j_0(kr) = \frac{3}{R^3}\int_0^{R} dr \; r^2 j_0(kr) \;,
\end{equation}
which can be used to compute the most likely value for the amplitude of the bulk flow \citep{Andersen:2016ywu}
\begin{equation}
   |\textbf{V}|=\frac{2}{3}\sigma_v(R)\;.  
\end{equation}
We are now in the position of highlighting the relation between $\mathcal{B}_R$ and the bulk flow amplitude $|\mathbf{V}|$. Substituting Eqs.~\eqref{psipar} in Eq.~\eqref{bulkinsphere} we can write
\begin{equation}
\mathcal{B}_R= \frac{3}{R^3} \frac{H^2f^2a^2}{2\pi^2}\int_0^\infty dk \int_0^{R} dr\; r^2 j_0(kr) P(k) = \frac{f^2H^2a^2}{2\pi^2}\int_0^{\infty} dk P(k) \tilde{W}(k,R)\;,
\end{equation}
which compared with Eq.~\eqref{sigmav} shows that the main difference between the two quantities is in the exponent of the window function $\tilde{W}$. The reason is that the standard definition of $\sigma_R$ (and $\sigma_v (R)$) quantifies the correlation between the smoothed density ( velocity) field at two points $\langle \delta_R(x),\delta_R(y) \rangle$, where the smoothed density field \begin{equation}
    \delta_R (x) = \int_0^\infty dy W(x-y) \delta(y)\;,
\end{equation}
is given by the convolution of $\delta$ with the top hat window function. Therefore, each $\delta_R$ contributes a Fourier transform of the window function in Eq.~\eqref{sigmav}. In contrast, Eq.~\eqref{bulkinsphere} is by definition the smoothed velocity correlation function, rather than the correlation function of the smoothed velocity field.\footnote{ {Notice that indeed a similar relation holds true between the rms of the density field $\sigma_R$ and the volume average of the correlation function $R^{-3} \int_0^R dr\; r^2 \xi(r)$, whose integrands in Fourier space differ only of a factor $\tilde{W}$.}}

To move forward, let us notice that on linear scales, if the velocity field is sourced by a homogeneous gaussian random density field (generating a scalar only gravitational potential), the parallel and perpendicular components of the velocity correlation functions are not independent \citep{gorski}
\begin{equation}\label{gorskicon}
\Psi_\parallel(r)= \frac{d}{dr} \left(r \Psi_\perp(r)\right) \;.   
\end{equation}
Simple integration by parts allow us to write
\begin{equation}
\int_0^R r^2 \Psi_\parallel(r)= \int_0^R r^2\frac{d}{dr}(r\Psi_\perp(r)) = R^3\Psi_\perp(R) - \int_0^R 2r^2\Psi_\perp(r)\;,  
\end{equation}
and conclude, by inserting this expression into Eq.~\eqref{bulkinsphere}, that a suitable expectation is $\bar{\mathcal{B}}_R = 3\Psi_\perp(R)\;$ (a bar indicates that this is the total velocity one would obtain \textit{if} the velocity field is irrotational and sourced by a scalar random density field, and may not be equal to what one would measure if the Universe is inhomogeneous, anisotropic or does not obey these assumptions). We can also define an analogous statistic $\mathcal{S}$ to the correlation dimension $D_2$ for $\mathcal{B}_R$:
\begin{equation}\label{Mateparam}
    \mathcal{S}(R) = \frac{1}{3}\frac{d}{dR}\left(R\mathcal{B}_R\right)\;,
\end{equation}
where we included a factor of $R$ in the parenthesis because, under the same assumptions as in Eq.~\eqref{gorskicon}, one can show that $\bar{S}=\Psi_\parallel$, which facilitates its physical interpretation.

\subsubsection*{Scaling of $\mathcal{B}_R$ and the transition to homogeneity}

{As mentioned above, while the specific threshold $\lambda$ used to define the homogeneity scale is arbitrary, it is the asymptotic behaviour of the volume average of the correlation function (and its scaling $D_2$) that determines whether a distribution is homogeneous. In this section, we show that this asymptotic behaviour is captured in the same way by the scaling of both $\mathcal{N}$ and $\mathcal{B}_R$. Indeed, apart from constant (or redshift-dependent) multiplicative amplitude factors, the radial dependence of $\mathcal{N}$ and $\mathcal{B}_R$ can be written as
\begin{equation} \label{I1}
\mathcal{N}(R) \propto I_1(R) = \int dk \;k^2 \;P(k)\,\tilde{W}(k,R) \;,    
\end{equation}
\begin{equation} \label{I2}
\mathcal{B}_R(R) \propto I_2(R) = \int dk\; P(k)\,\tilde{W}(k,R) \;.     
\end{equation}
For a fiducial $\Lambda$CDM power spectrum $P(k)$ computed using CLASS \citep{2011arXiv1104.2932L}, the ratio of the logarithmic derivatives of $I_1$ and $I_2$ is shown in Fig.~\ref{RatioD/S} before (orange line) and after (green line) subtraction of the BAO peak.
\begin{figure}[h]
\centering
\includegraphics[scale=0.8]{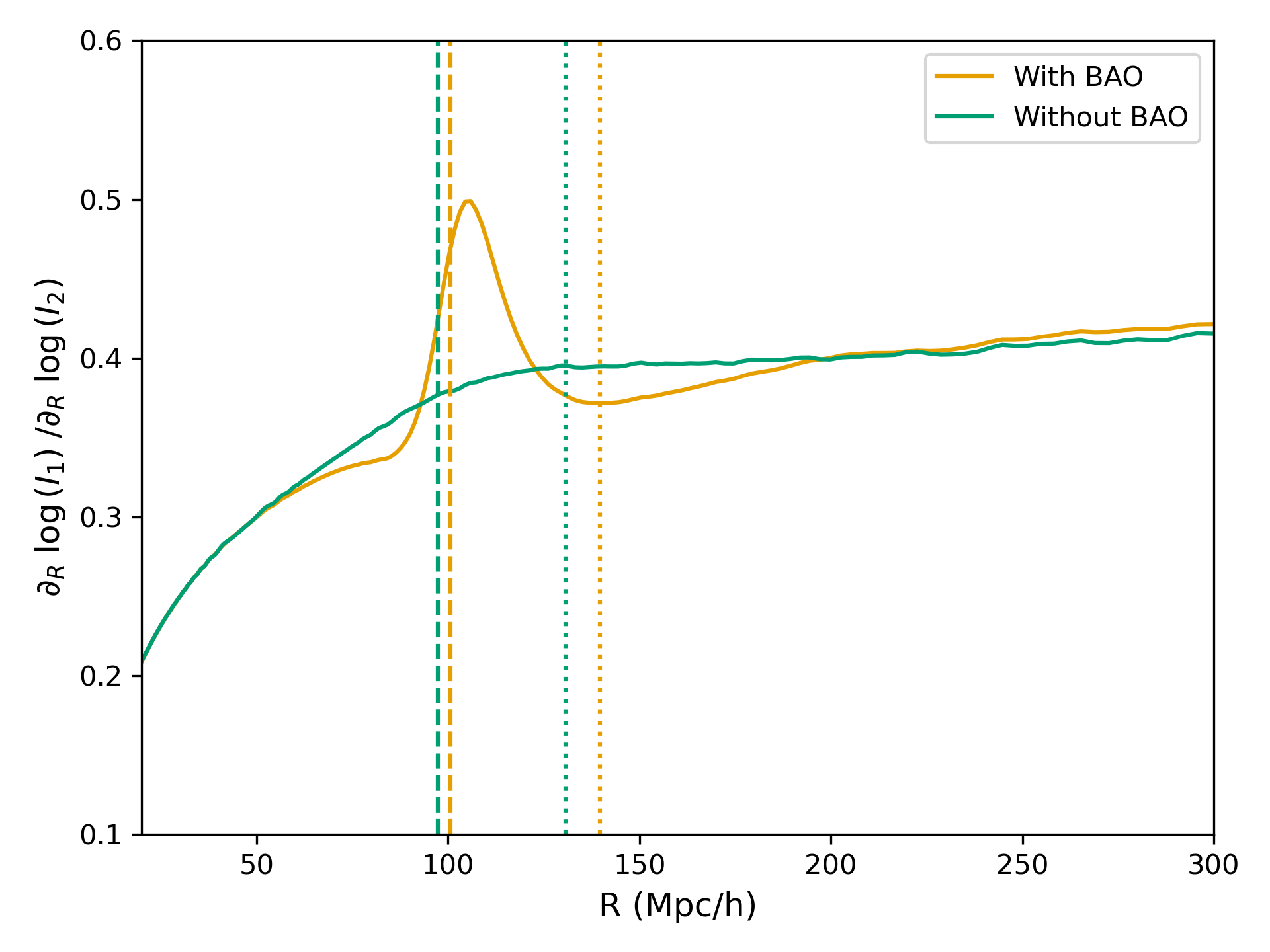}
\caption{{The ratio of the logarithmic derivatives of the integrals in Eqs.~\eqref{I1} and \eqref{I2}, capturing the asymptotic behaviour of the proportional scalings of $\mathcal{N}$ and $\mathcal{B}_R$. The green solid curve is obtained by removing the BAO feature through Gaussian smoothing of the $P(k)$ wiggles. The dashed vertical lines correspond to the velocity coherence scales defined by Eq.~\eqref{velcoscale}, whereas the dotted vertical lines correspond to the density homogeneity scale $R_\rho$. In both cases, at sufficiently large scales the ratio approaches a constant value, highlighting the shared asymptotic scaling of the estimators $\mathcal{N}$ and $\mathcal{B}_R$.}}
\label{RatioD/S}
\end{figure}
Remarkably, the ratio as a function of $R$ approaches a constant value, indicating that on sufficiently large scales $\mathcal{B}_R$ and $\mathcal{N}$ share the same asymptotic behaviour. In the figure, the dashed lines correspond to the scale at which the product $R\mathcal{B}_R$ reach its maximum. Interestingly, the numerical value of this scale does not change significantly whether the BAO feature is subtracted or not.

\subsection{The velocity coherence scale}\label{velcohscalsec}

{The evolution of $\mathcal{B}_R$ as a function of radius exhibits a very interesting feature. As one might naively expect from its close correspondence with the bulk flow amplitude, it is a decreasing function of the sphere radius. However, on small scales, larger fluctuations of the density field induce strongly correlated motions, causing $\mathcal{B}_R$ to decrease at a slower rate than on large scales. In particular, as shown in Fig.~\ref{fig:hom_scale}, $\mathcal{B}_R$ decreases more slowly than $R^{-1}$ below a certain scale, corresponding to a maximum in $R\mathcal{B}_R$, and faster beyond it. This maximum corresponds to a zero of $\mathcal{S}$, the derivative of $R\mathcal{B}_R$, and it is what we define as the \textit{velocity coherence scale} $R_v$, satisfying
\begin{equation}\label{velcoscale}
    \mathcal{S}(R_v) = 0 \;, \qquad \text{sup}(R\mathcal{B}_R)_{\forall R}= R_v\mathcal{B}_R(R_v)\;.
\end{equation}
}

{As shown in Fig.~\ref{RatioD/S}, the velocity coherence scale can be used as a pivot scale to identify the sphere size at which $\mathcal{N}$ and $\mathcal{B}_R$ transition to the same asymptotic scaling. Since it is precisely the asymptotic scaling of the volume-averaged correlation function that defines the homogeneity of a matter distribution, we advocate that $R_v$, just like $R_\rho$, can be used to estimate the transition to cosmic homogeneity.}

\begin{figure}[h]
    \centering
\hspace*{-1cm}\includegraphics[scale=0.55]{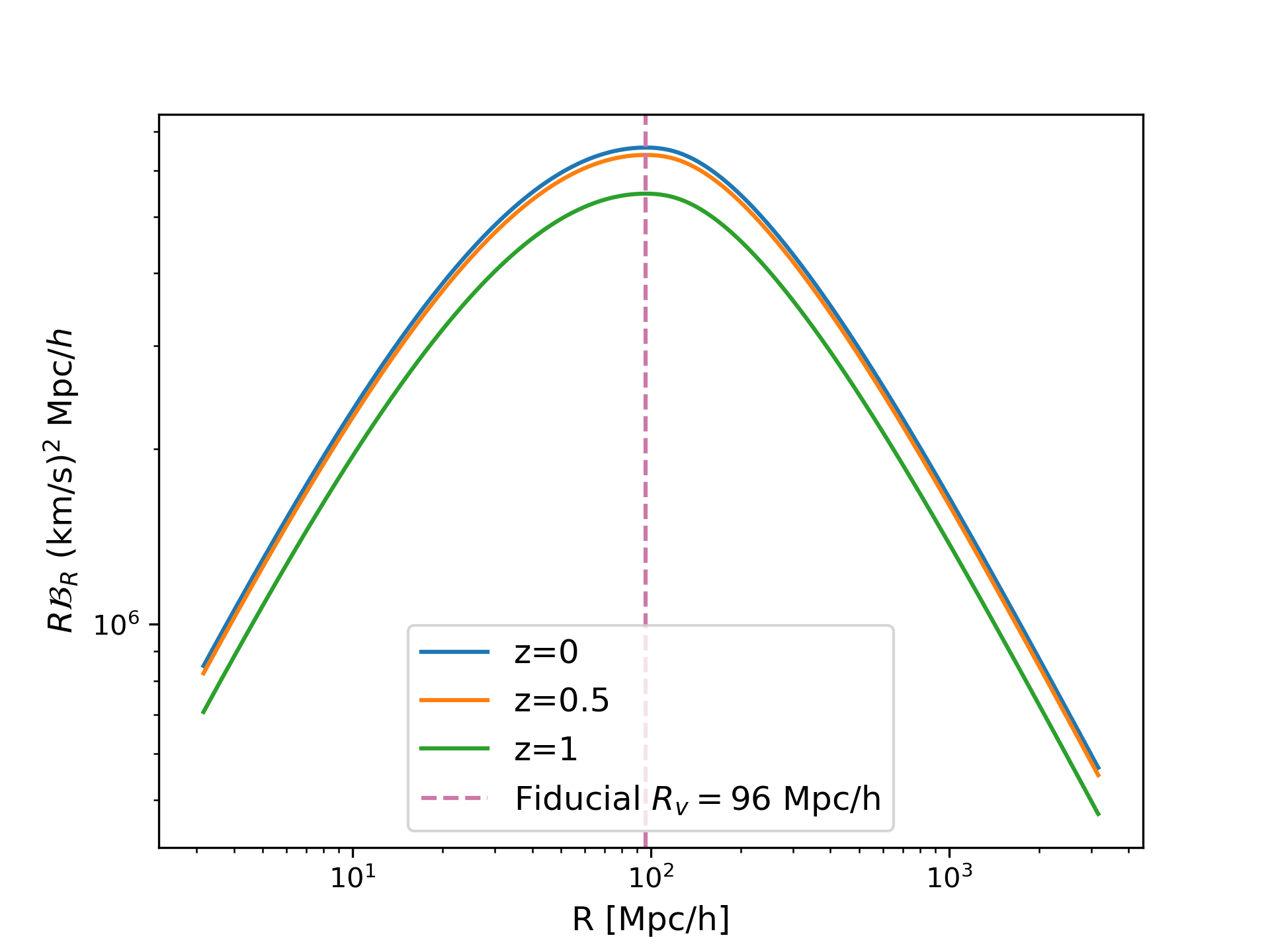}\;\;\includegraphics[scale=0.55]{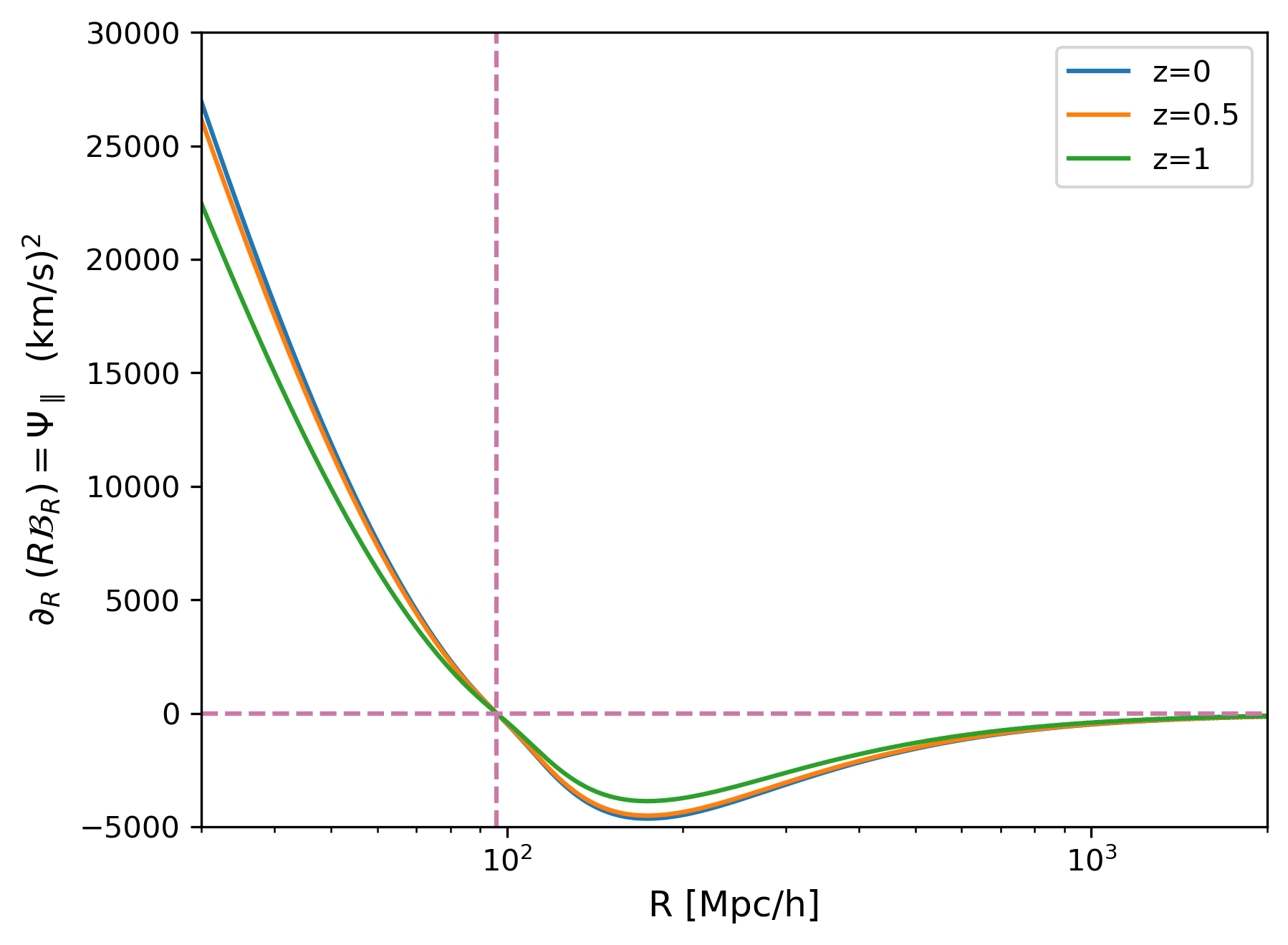}
    \caption{The function $R\mathcal{B}_R$ and its derivative for a fiducial $\Lambda$CDM cosmology. We identify the homogeneity scale with the transition from correlated to anticorrelated averaged velocities along the vector separation of galaxy pairs ($\Psi_\parallel=0$), corresponding to a change in slope in the radial evolution of $\mathcal{B}_R$ and the smoothed velocity variance $\sigma_v$.}
    \label{fig:hom_scale}
\end{figure}

{Notice that since in linear theory $\mathcal{S}=\Psi_\parallel$, the identification of the first zero of $\mathcal{S}$ with the scale of transition to homogeneity has a straightforward physical interpretation: the zero crossing of $\Psi_\parallel$ signals that no external potential gradient is sourcing coherent flows throughout the sphere (see Fig.~\ref{diagram} for a diagrammatic representation).}

{We are now left with two statistics, $\mathcal{B}_R$ and $\mathcal{S}$, and a theoretical prediction relating them in a homogeneous and isotropic universe to $\Psi_\parallel$ and $\Psi_\perp$, which are themselves connected by Eq.~\eqref{gorskicon}.}

\begin{figure}
    \centering
    \includegraphics[scale=0.5]{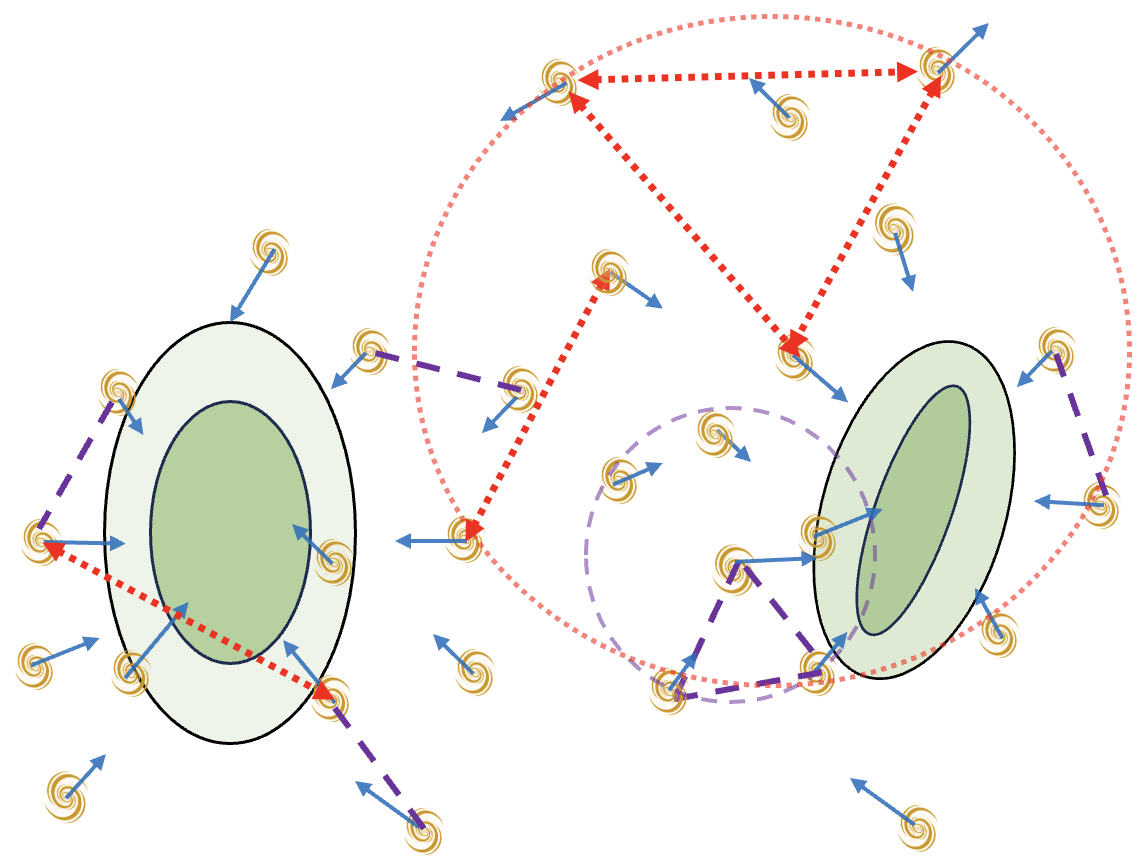}
    \caption{Acting on small scales (purple dashed lines), density gradients on large scales source collective flows towards a common attractor (green ellipses), resulting in a positive correlation ($\Psi_\parallel > 0$). The lack of strong density perturbations on large scales, on the other hand, causes galaxies to move in opposite directions rather than together (red dotted lines), contributing a negative $\Psi_\parallel$. The homogeneity scale corresponds to the scale at which $\Psi_\parallel=0$, i.e. where galaxy motions with respect to each other are, on average, uncorrelated.}
    \label{diagram}
\end{figure}

\subsubsection*{$R_v$ sensitivity to long-range correlations}

{A key feature of the peculiar velocity (PV) field is that it is much more sensitive to large-scale density modes than to small-scale ones, reflecting the $k^{-1}$ suppression arising from the Euler equation. For this reason, we expect $\mathcal{B}_R$ and $\mathcal{S}$ to be much more sensitive to large-scale correlations than $\mathcal{N}$ and $D_2$. To illustrate that this is indeed the case, we consider a set of toy models for the power spectrum $P(k)$ in which the Harrison–Zeldovich behaviour at $k\leq k_{eq}$, where $k_{eq}=a_{eq}H_{eq}$ is the scale at which a mode entered the horizon at matter–radiation equality, is replaced with power laws $P(k)\propto k^n$ with $n\neq n_s$. We also include a toy model with $n=-1$, which is still homogeneous according to the definition given in Appendix~\ref{appendixA}, but for which $\int\xi(r)\rightarrow\infty$ as $r\rightarrow \infty$. This behaviour, as argued in Ref.~\cite{Gabrielli:2001xw}, is typical of thermodynamical systems at the critical point of a second-order phase transition, which exhibit large-scale critical correlations.}

{We plot the $P(k)$ and the corresponding correlation functions $\xi(r)$ in Fig.~\ref{toymodels}, explicitly showing the emergence of large-scale correlations in the $n=-1$ case (blue line).}

\begin{figure}[h]
\centering
\hspace*{-0.8 cm}
\includegraphics[scale=0.53]{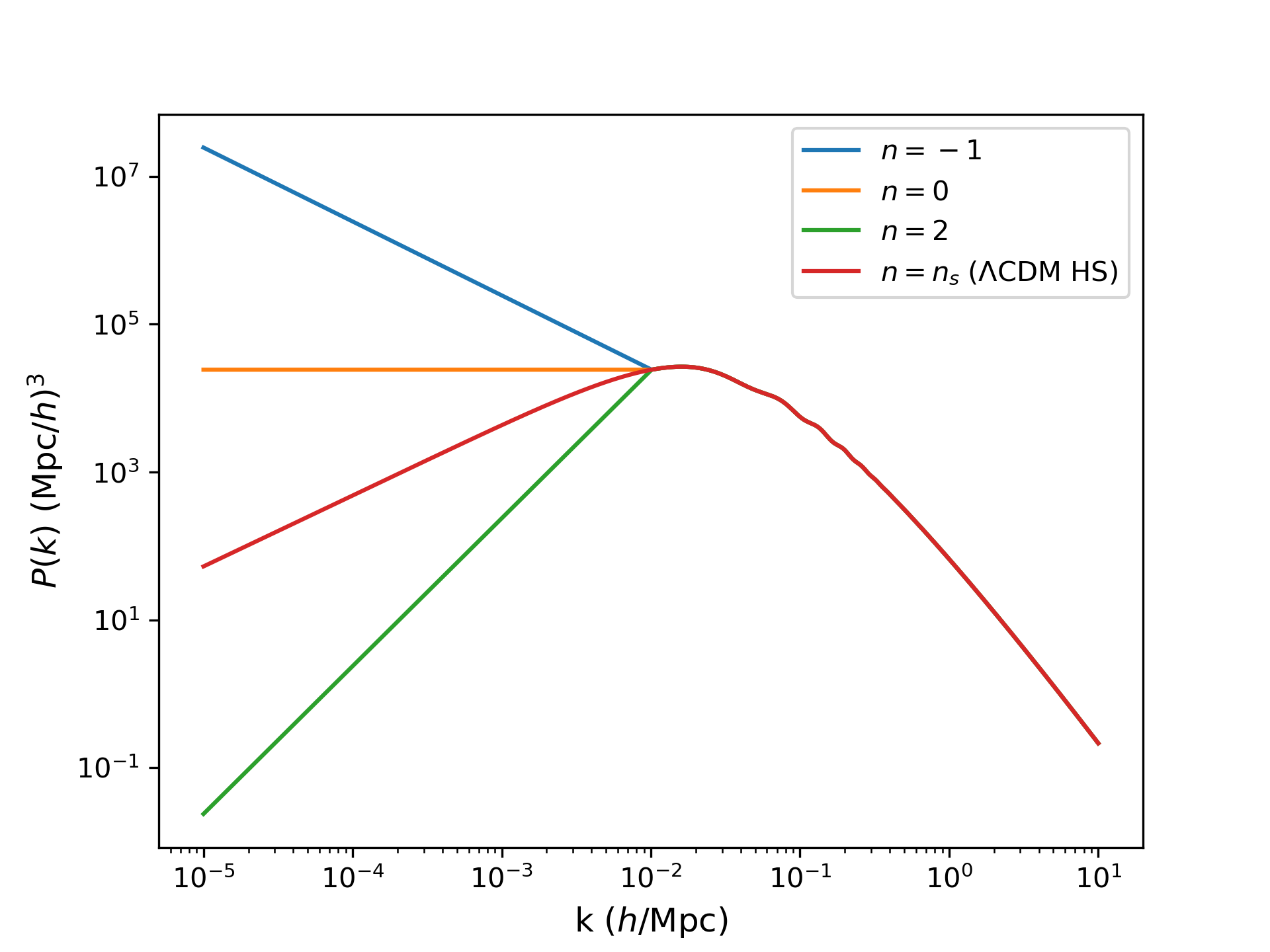}\includegraphics[scale=0.53]{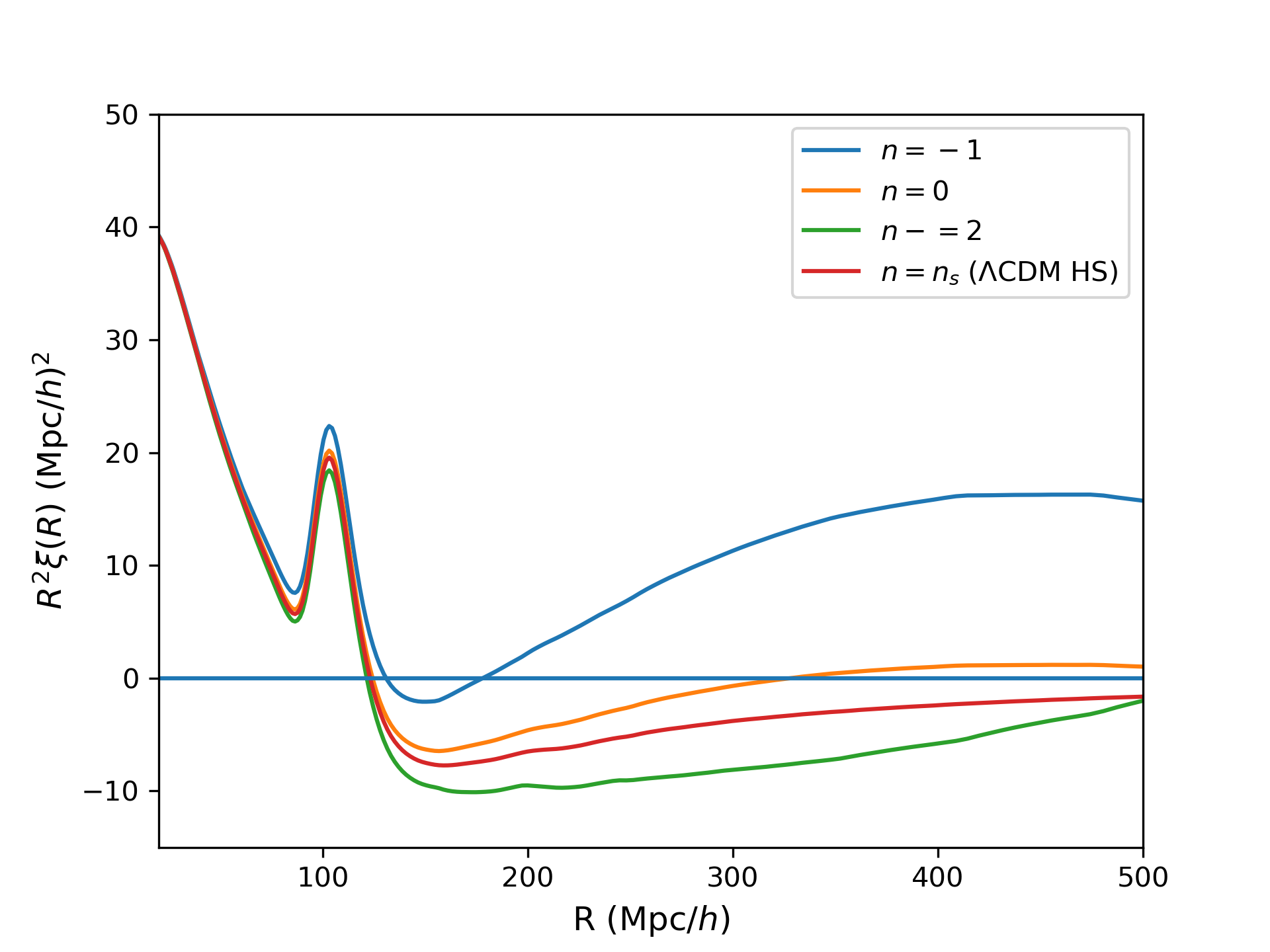}
    \caption{{The toy models of $P(k)$ and their corresponding correlation functions $\xi(r)$ used to compare distributions with different homogeneity properties.}}
    \label{toymodels}
\end{figure}

{In Fig.~\ref{DVtoymodels} we compare the correlation dimension $D_2$ and $\mathcal{S}$ computed for these toy models, together with the corresponding thresholds for $R_\rho$ and $R_v$. Whilst all the distributions have well-behaved asymptotic values of $D_2 \rightarrow 3$, we notice that for the case $n=-1$ a velocity coherence scale is not defined, since $\mathcal{S}$ never vanishes.\footnote{In hindsight, this is easily explained by the lack of a $k^2$ scaling in the integral of $P(k)$ entering the velocity correlation functions.} We conclude that whilst the existence of a velocity coherence scale ensures the existence of a homogeneity scale, the opposite is not necessarily true. In particular, for $R_v$ to exist one cannot have long-range fluctuations (i.e. super-Poissonian), and at large scales we must have $P(k) \propto k^n$ with $n\geq0$, corresponding to at least Poissonian noise or sub-Poissonian (superhomogeneous) fluctuations according to the terminology of Ref.~\cite{Gabrielli:2001xw}.}

{This simple exercise highlights the differences between $R_v$ and $R_\rho$, showing that the former, beyond providing complementary information on the transition to homogeneity, is particularly sensitive to large-scale correlations in the density field that are forbidden in the standard cosmological model and its Harrison–Zeldovich power spectrum of primordial perturbations. Furthermore, as we show explicitly in Appendix~\ref{appendixB} by directly comparing the values of $R_v$ and $R_\rho$ computed across different cosmologies, the $1\%$ threshold used to define $R_\rho$ makes its numerical value quite susceptible to the BAO scale and amplitude. This occurs because minimal changes in the cosmological parameters can shift the flattening of the $D_2$ curve after the BAO peak just above or below the $2.97$ threshold, leading in extreme cases to $\approx 20\%$ variations in $R_\rho$ induced by less than a $1.5\%$ variation in $H_0$, (despite the identical asymptotic behaviour of the correlation functions). In turn, this makes the comparison of the homogeneity scale across different datasets prone to biases if, for example, their fiducial $H_0$ values are not in perfect agreement. In contrast, as we will show later, the velocity coherence scale $R_v$ is mostly sensitive to $k_{eq}$ and varies continuously with it. Furthermore, standard measurements of the velocity correlation functions are based on log-distance ratios $\eta$, for which changes in $H_0$ are usually absorbed into the zero-point calibration of the distance indicator. Consequently, varying the Hubble constant primarily amounts to relabelling distances between Mpc and $h^{-1}\mathrm{Mpc}$, leaving measurements of the $\Psi$ functions in these respective units virtually unchanged.

\begin{figure}[h]
\begin{center}   
\hspace*{-1.2 cm}
\includegraphics[scale=0.55]{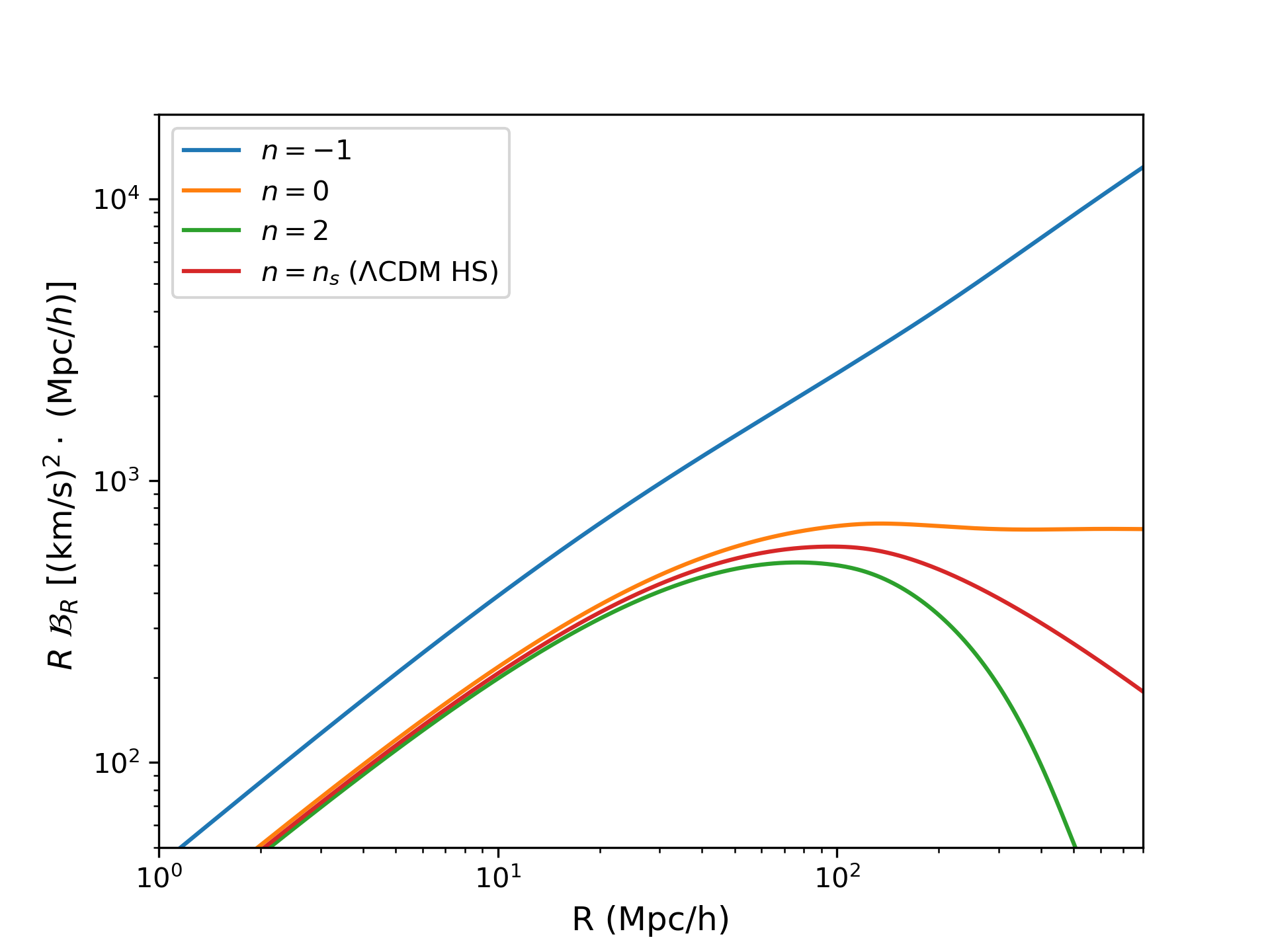}\includegraphics[scale=0.55]{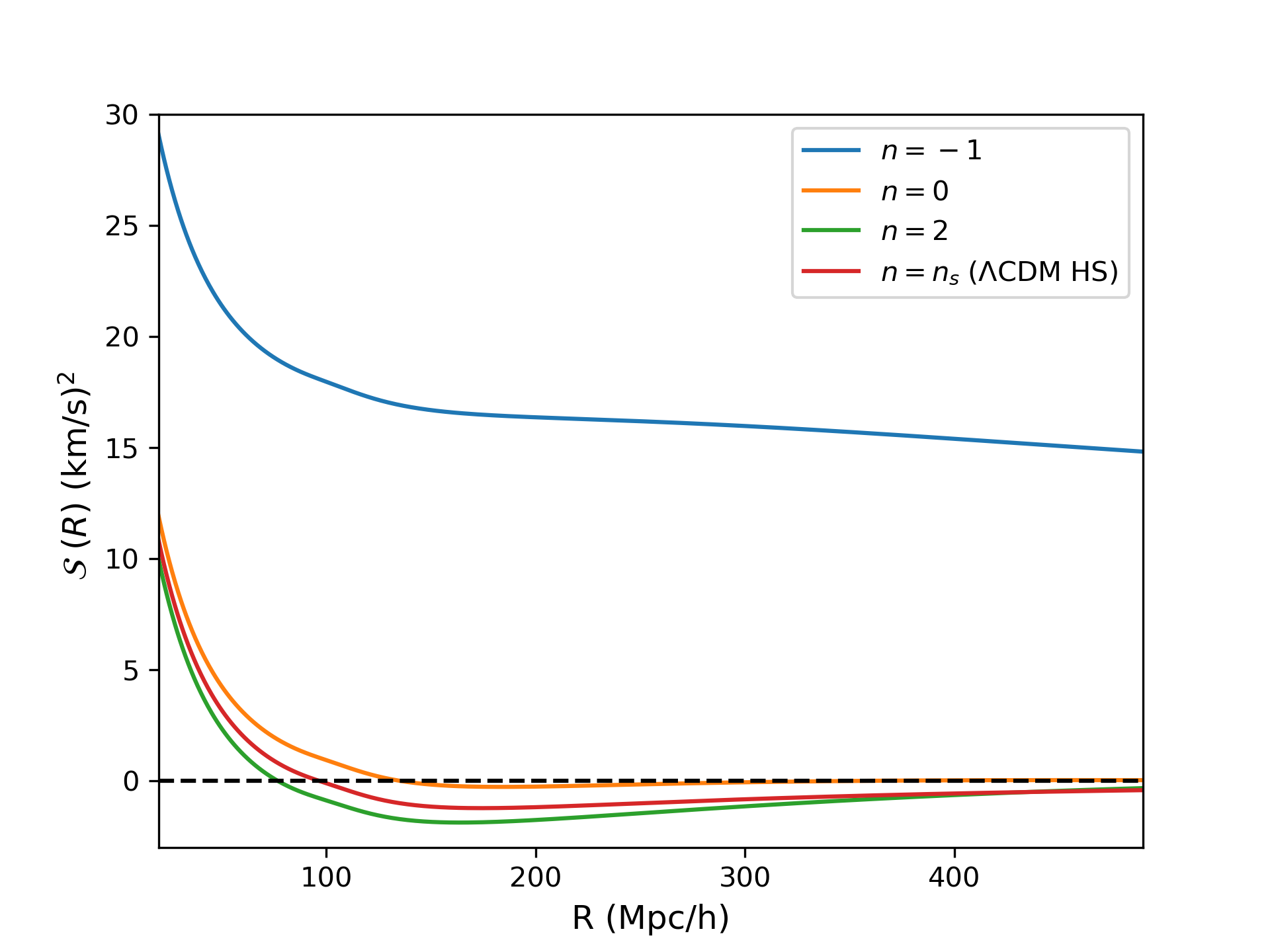}\\
\includegraphics[scale=0.55]{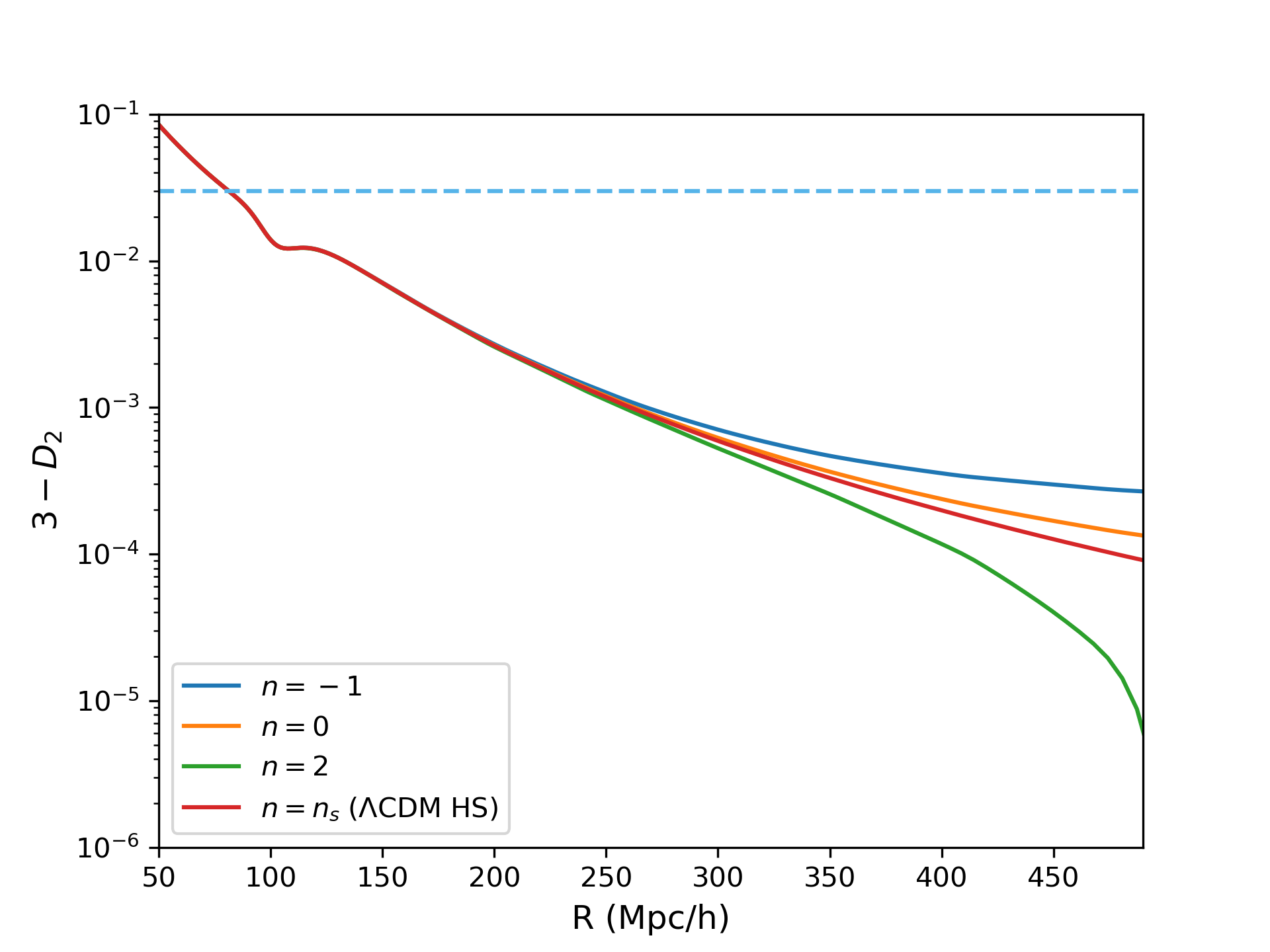}
\end{center}
    \caption{{\textit{Upper panels:} evolution of $R\mathcal{B}_R$ and $\mathcal{S}$ as a function of distance for the toy models of $P(k)$ shown in Fig.~\ref{toymodels}. The lack of a well-defined maximum in the top-left panel and of a zero-crossing in the top-right panel for the case $n=-1$ highlights that $R_v$ only exists when the volume integral of the correlation function does not diverge as $R \rightarrow \infty$. \textit{Bottom panel:} the homogeneity scale $R_\rho$, identified with the intersection of the curves with the threshold $D_2 =2.97$, is essentially the same for all the models considered. Nevertheless, differences in the asymptotic behaviour of the curves grow with the radius $R$.}}  
    \label{DVtoymodels}
\end{figure}

\subsection{$R_v$ and its potential use as a standard ruler}

{Fig.~\ref{fig:hom_scale}} shows an interesting feature of the velocity coherence scale: its value in comoving coordinates ($\rm{Mpc}/h$) is redshift independent. This is easily explained by the fact that in linear theory the time evolution of the velocity correlation functions is entirely encoded in the amplitude factors multiplying the integrals in Eqs.~\eqref{psipar}. Therefore, whilst the overall amplitude of the correlation functions evolves with time, the location of the maximum in $R\mathcal{B}_R$ remains constant. In Appendix~\ref{appendixC} we show, using ``triangular'' toy models of $P(k)$, that the value of $R_v$ is essentially determined by the value of $k_{eq}$, or equivalently by $\omega_m$ and the baryon fraction. Therefore, the velocity coherence scale can serve as a standard ruler in the same way as the turnover scale of the matter power spectrum \citep{2013MNRAS.429.1902P}. That the velocity field carries information about this scale has been noted, for example, in Ref.~\cite{Lai:2025xxg}.}

{Inspired by the form of the analytical solutions obtained for the toy models in Eqs.~\eqref{rvanal}, we postulate the following analytical relation between $R_v$ and $k_{eq}$:
\begin{equation}\label{rvofkeq}
 R_v(k_{eq})\approx \frac{\alpha}{k_{eq}} + \beta k_{eq} + \gamma\;.  
\end{equation}}

{We fit this expression numerically to the values of $R_v$ computed using CLASS, varying $\omega_m$ in the range $0.01 \leq \omega_m \leq 0.3$ while fixing the present-day matter density to $\Omega_m=0.31$ and the baryon fraction to $f_b=0.156$, and keeping the remaining parameters consistent with the Planck 2018 best-fit cosmology \citep{Planck:2018vyg}. We obtain the best-fit coefficients
\begin{equation}
    \alpha\approx 0.108\;, \qquad \beta\approx -1960\;,\qquad \gamma\approx 101\;,
\end{equation}
with the best fit and the residuals shown in Fig.~\ref{rvvskeq}. The figure also shows the numerical evaluation of $R_v$ obtained from non-linear computations of $P(k)$ in CLASS using \texttt{halofit}. As shown by the similar residuals, the numerical fit for $R_v(k_{eq})$ remains satisfactory (with  maximum deviations from the CLASS output below 3 $\%$) even in the presence of non-linear modes. }

\begin{figure}[h]
    \centering
    \includegraphics[scale=0.8]{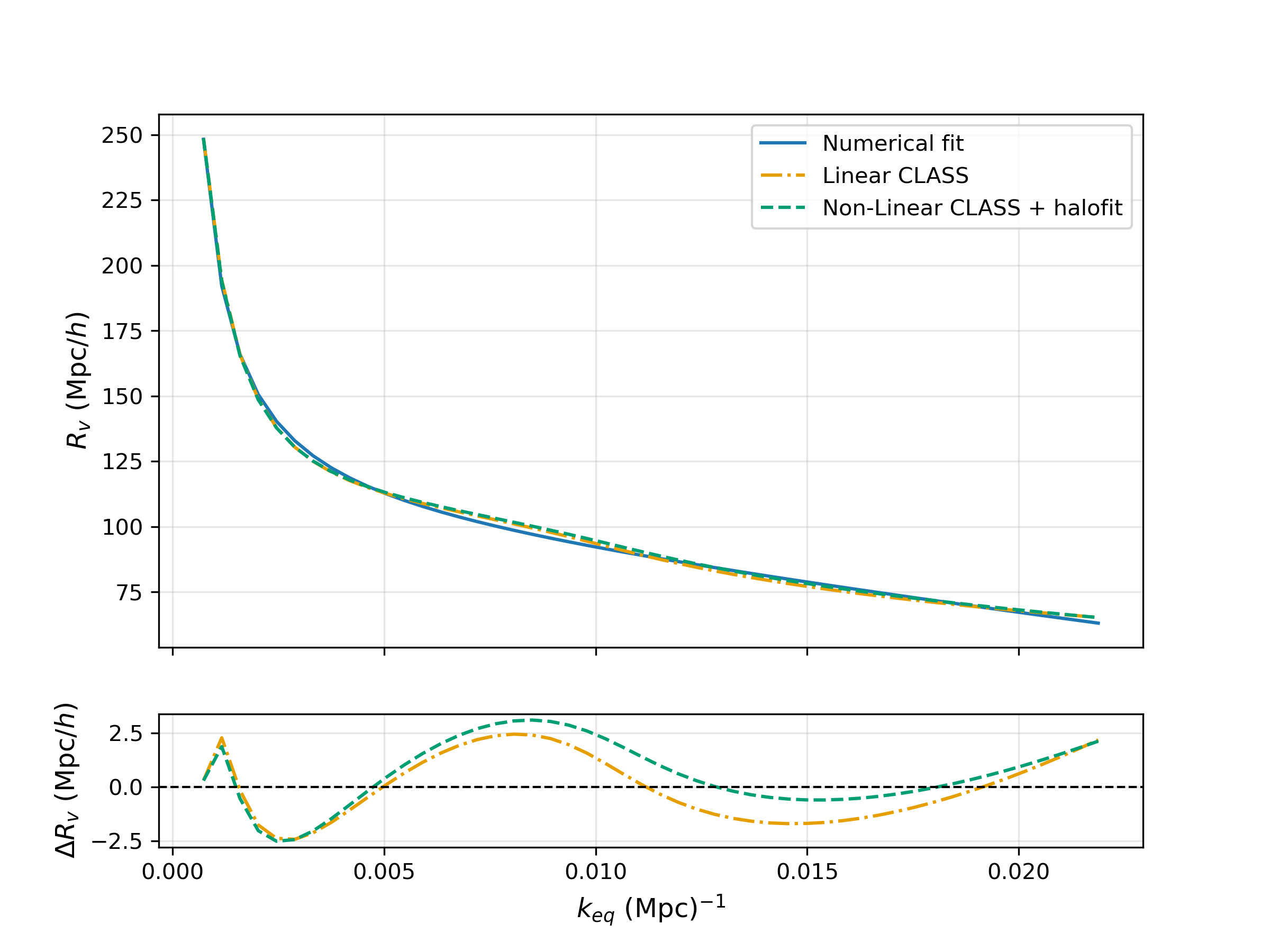}
    \caption{{The value of $R_v$ as a function of $k_{eq}$ obtained while keeping fixed the present-day matter density $\Omega_m$ and the baryon fraction $f_b$, using both the linear (dot-dashed orange line) and non-linear (dashed green line) $P(k)$. The solid blue line shows the best-fit prediction of the parametric model in Eq.~\eqref{rvofkeq} to the linear $P(k)$. The bottom panel shows that the residuals in both cases are small and remain below $2\%$ across the entire $k_{eq}$ range.}}
    \label{rvvskeq}
\end{figure}

%%%%%%%%%%%%%%%%%%%%%%%%%%%%%%%%%%%%%%%%%%%%%

\section{Proof of Concept on SDSS data}\label{SDSS}
To assess {whether measurements of $R_v$ are within the reach of current cosmological surveys}, we attempt to infer it from measurements of the velocity correlation functions obtained in \cite{Lyall:2024mks} from the Sloan Digital Sky Survey (SDSS) and its companion suite of  2048 mocks \citep{Howlett:2022len}, which is the largest homogeneously-selected publicly-available PV catalogue. The data consists of $\approx 34,000$ measurements of {distances of early type elliptical galaxies obtained using the fundamental plane empirical relation, with a mean distance error across the dataset of $\approx 23\%$. After subtracting the homogeneous Hubble flow, these distances are converted in measurements of} PVs over a sky area of $7000$ deg$^2$ up to a redshift $z=0.1$. These observations were compressed into summary statistics comprising 25 equally spaced binned measurements of $\Psi_\parallel(r)$ and $\Psi_\perp(r)$ up to a distance of $150 \rm{Mpc/h}$. 

Fig.~\ref{scaling} shows these measurements for the data, a fiducial $\Lambda$CDM cosmology, the individual mocks and their mean.
\begin{figure}[t]
\begin{center}
    
\hspace*{-1cm}
\includegraphics[scale=0.7]{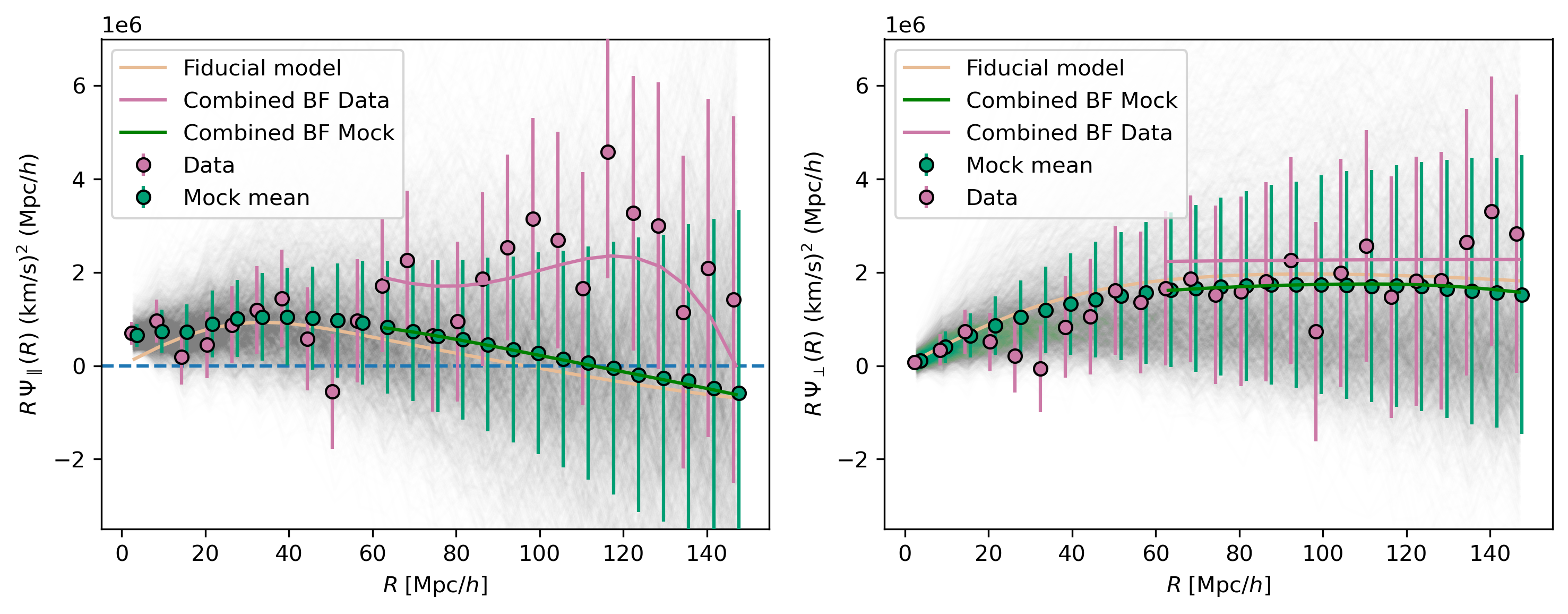} 
    \caption{ The functions $\Psi_\parallel$ and $\Psi_\perp$ for the SDSS data (pink), each mock realization (faint solid gray lines) together with their mean (green dots) and the fiducial cosmology used to build the mocks (orange). The solid pink and green lines correspond to the best fit results from our MCMC analysis (see Figs.~\ref{mock_validation},\ref{data_results} and Sec.~\ref{SDSS}). One can easily see that whilst the mean of the mocks recovers very well the input fiducial model, the distribution of the mocks around this mean is skewed and non-gaussian.   
}
    \label{scaling}
\end{center}
\end{figure}

{We identified $R_v$ with} the turnover point in $R \mathcal{B}_R$, which as we shown is proportional to $R\Psi_\perp (R)$ using Eq.~\eqref{gorskicon}. Consistency also requires that the same scale corresponds to the zero crossing of $\Psi_\parallel(R)$. As such, we look for the peak in $R\mathcal{B}_R$ adopting the same methodology used in searches of the turnover scale of the matter power spectrum, see e.g. \cite{Blake:2004tr}, by fitting $R\Psi_\perp$ using a piece-wise parabolic function
\begin{equation}\label{turnaround}
R \mathcal{B}_R^{\rm{model}}= 3R\bar{\Psi}_\perp=\begin{cases}
    C\left(1-\alpha x^2\right) \; \textrm{if} \;\;R < R_v \\
    C\left(1-\beta x^2\right) \; \textrm{if} \;\;R \geq R_v
\end{cases} \qquad x=\left(R - R_v\right)/R_v\;.
\end{equation}
At the same time, we look for the zero crossing of $\Psi_\parallel$ by modeling it with a third order polynomial
\begin{equation}\label{model2}
    \Psi_\parallel= c_0 + c_1 R + c_2 R^2 + c_3 R^3\;,
\end{equation}
noticing that the zero crossing condition at the homogeneity scale fixes the coefficient $c_0$ to be
\begin{equation}
    c_0= -(c_1 R_v + c_2R_v^2 + c_3R_v^3)\;.
\end{equation}
The choice of a third order polynomial has no physical motivation, but introduces the same number of free parameters as in the parabolic model for $R\Psi_\perp$, allowing for a fairer comparison of the two estimators. We also verified, see appendix \ref{appendixD}, that increasing the number of free parameters does not improve appreciably the fit.

We perform an MCMC exploration of the parameter space for the variables $C,\alpha,\beta,c_1,c_2,c_3$ and $R_H$ using the \textit{emcee} ensamble sampler, and \textit{ChainConsumer} to analyse our chains \citep{2013PASP..125..306F,Hinton2016}.\footnote{Available at https://emcee.readthedocs.io/en/stable/ and https://samreay.github.io/ChainConsumer/} We use a Gaussian likelihood
\begin{equation}\label{likelihood}
    \log\mathcal{L}_{\rm{Gaussian}}=\sum_i -\frac{1}{2}\left[ \left(d^i-m^i\right)^T {\rm Cov}^{-1}_{ij}\left(d^j -m^j\right) \right]\;,
    \end{equation}
where the subscripts $i,j$ runs over the binned measurements, and where the data-vector $d_i$ is constructed by stacking measurements of $R\Psi_\perp(R)$ and $\Psi_\perp(R)$,  ${\rm Cov}_{ij}$ being their covariance computed from the 2048 mock realisations, and where the model predictions $m_i$ are obtained from Eqs.~\eqref{turnaround} and \eqref{model2}. To avoid contamination from small-scale non-linearities in the zero-crossing and turnaround fitting, we restrict our analysis to measurements above $60 ~\mathrm{Mpc}/h$. Since $R\mathcal{B}_R$ is positive, following the same prescription used for $P(k)$ in \cite{Blake:2004tr}, we must require $\alpha,\beta \leq 1$. We adopt uniform priors $-1 \leq\left[\alpha, \beta\right]\leq 1$,  $10^5\leq C \leq 10^8$, $ -50000< c_1 <50000$, $-100 \leq \left[c_2,c_3 \right]\leq 100$ and $10\leq R_v \leq 200$.\footnote{To assess the convergence of the chains we follow the prescription given in \href{https://emcee.readthedocs.io/en/stable/user/autocorr/}{https://emcee.readthedocs.io/en/stable/user/autocorr/} and check the estimated autocorrelation time $\tau$ every 100 steps for each
chain, considering it convergent if the estimate has changed by less then 1\%.}
Fig.~\ref{mock_validation} shows the results of our analysis for the mocks, fitting jointly or separately (with the appropriate subsets of the full covariance matrix) the mock mean using the turnaround and polynomial models for $\Psi_\parallel$ and $R\Psi_\perp$. {The resulting $R_v\approx 105 \substack{+45 \\ -38}~ \mathrm{Mpc}/h$ is compatible with the value computed for the fiducial $\Lambda$CDM cosmology used to produce the 2048 mocks, $R_v\approx 96 \mathrm{Mpc}/h$.
} Fig.~\ref{data_results} shows instead the results of our pipeline for the SDSS data. 
\begin{figure}[htbp]
\centering
\includegraphics[scale=0.42]{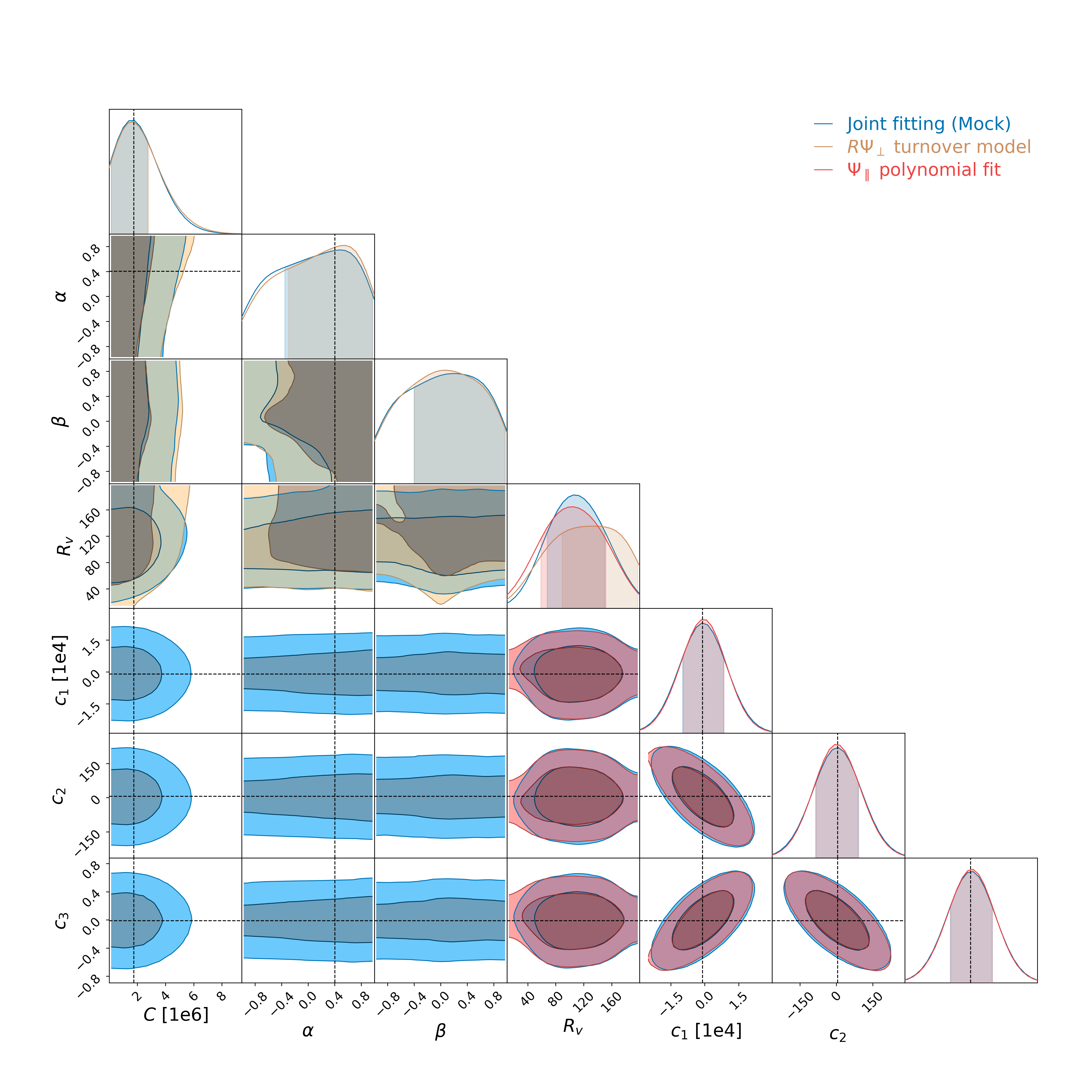}
    \caption{The results of our MCMC explorations for the models defined in Eqs.~\eqref{turnaround},\eqref{model2} fitted to the mock mean separately (orange and red) or jointly (blue).}
    \label{mock_validation}
\end{figure}

\begin{figure}[htbp]
\centering
\hspace*{-1cm}
\includegraphics[scale=0.42]{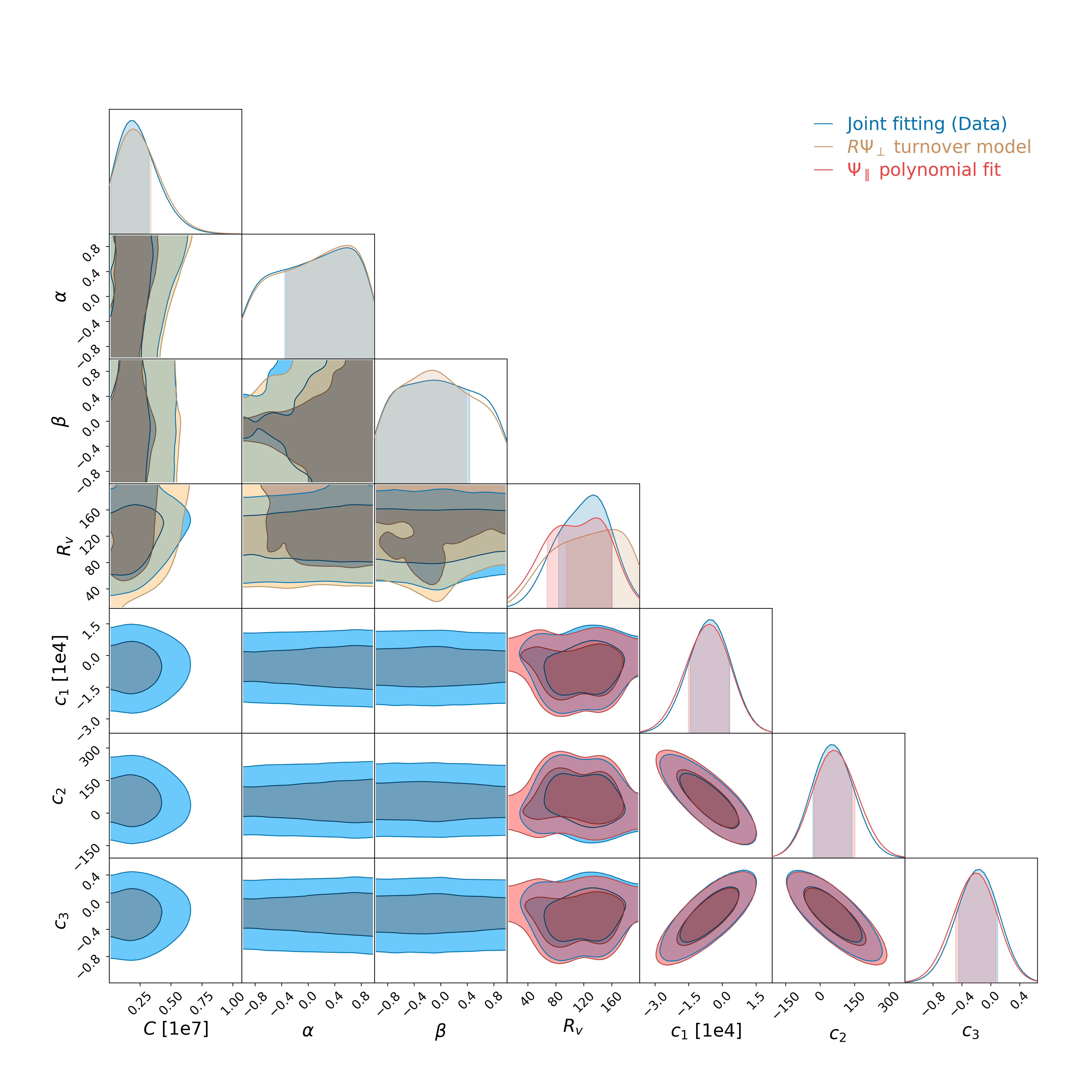}
    \caption{As in Fig.~\ref{mock_validation}, but for the SDSS data. Compared to the results from the mocks, the parameters of the polynomial fitting of $\Psi_\parallel$ are much less constrained. This is non-suprising given the noticeable difference between the green and pink lines in the left panel of Fig.~\ref{scaling}.  }
    \label{data_results}
\end{figure}

We found that both the mock mean and the data provide consistent marginalised constraints on the velocity coherence scale $R_v$, both from the combined analysis and fitting individually $\Psi_\parallel$ and $\Psi_\perp$. Table~\ref{tab:model_params} contains a summary of the results of our MCMC exploration. 
\begin{table}[t]
    \centering
    \renewcommand{\arraystretch}{1.25}
    \setlength{\tabcolsep}{5pt}
    \resizebox{\textwidth}{!}{%
    \begin{tabular}{lccccccc}
        \hline\hline
        Model
        & $C$
        & $\alpha$
        & $\beta$
        & $R_v \;(\mathrm{Mpc}/h)$
        & $c_1$
        & $c_2$
        & $c_3$ \\
        \hline

        \multicolumn{8}{c}{\textbf{Mock}} \\
        \hline
        Joint fit
        & $\left(10.0^{+17.2}_{-8.5}\right)\times10^{5}$
        & $1.0^{+0.0}_{-1.3}$
        & $0.17^{+0.80}_{-0.57}$
        & $105^{+45}_{-38}$
        & $\left(-0.5^{+8.9}_{-9.3}\right)\times10^{3}$
        & $8^{+84}_{-92}$
        & $0.02^{+0.26}_{-0.31}$ \\

        $R\Psi_{\perp}$ turnover model
        & $\left(1.2^{+1.6}_{-1.1}\right)\times10^{6}$
        & $1.0^{+0.0}_{-1.3}$
        & $0.06^{+0.91}_{-0.46}$
        & $\geq 90$
        & \textemdash
        & \textemdash
        & \textemdash \\

        $\Psi_{\parallel}$ polynomial fit
        & \textemdash
        & \textemdash
        & \textemdash
        & $100^{+52}_{-41}$
        & $\left(-0.2^{+8.7}_{-9.3}\right)\times10^{3}$
        & $-7^{+97}_{-79}$
        & $-0.01^{+0.29}_{-0.27}$ \\

        \hline
        \multicolumn{8}{c}{\textbf{Data}} \\
        \hline
        Joint fit
        & $\left(1.8^{+1.4}_{-1.7}\right)\times10^{6}$
        & $0.893^{+0.085}_{-1.245}$
        & $-0.08^{+0.52}_{-0.89}$
        & $138^{+23}_{-54}$
        & $\left(-5.5^{+8.9}_{-8.7}\right)\times10^{3}$
        & $51^{+87}_{-83}$
        & $-0.15^{+0.25}_{-0.29}$ \\

        $R\Psi_{\perp}$ turnover model
        & $\left(1.8^{+1.6}_{-1.7}\right)\times10^{6}$
        & $0.893^{+0.085}_{-1.222}$
        & $-0.13^{+0.51}_{-0.85}$
        & $\geq 100$
        & \textemdash
        & \textemdash
        & \textemdash \\

        $\Psi_{\parallel}$ polynomial fit
        & \textemdash
        & \textemdash
        & \textemdash
        & $141^{+18}_{-74}$
        & $\left(-5.6^{+8.7}_{-9.6}\right)\times10^{3}$
        & $55^{+96}_{-82}$
        & $-0.18^{+0.25}_{-0.31}$ \\

        \hline\hline
    \end{tabular}%
    }
        \caption{Summary of the marginalized constraints on the fitting parameters and the velocity coherence scale from our MCMC exploration, both for the combined likelihood and the individual fit to $\Psi_\parallel$ and $\Psi_\perp$.}
         \label{tab:model_params}
\end{table}
 Given the quite large $\geq 30\%$ uncertainty on the measurement, one could ask whether the accuracy of this method will improve with upcoming PV datasets. To answer, we apply the same analysis to a set of measurements centered on the mock mean, but artificially decreasing their uncertainties by decreasing the covariance matrix by a factor 5. Fig.~\ref{fake_err_bar} shows a comparison of the results. Whilst the constraints on the polynomial-fit parameters $c_i$ improve substantially, the uncertainties on the turnover fit and on the inferred homogeneity scale change only marginally. We argue that this is a consequence of the weak constraints on the parameter describing the parabolic downfall $\beta$. In hindsight, this is expected since SDSS measurements of the $\Psi$'s extend only up to $150~ \mathrm{Mpc}/h$, and the best fits indicate an homogeneity scale close to this upper bound. We conclude that to improve significantly the constraints on $R_v$, one would also need measurements of the $\Psi$'s at larger scales --- and that doing this even for current data may result in improved constraints on {$R_v$} compared to those presented here.
\begin{figure}[t]
\hspace*{-1cm}
\includegraphics[scale=0.42]{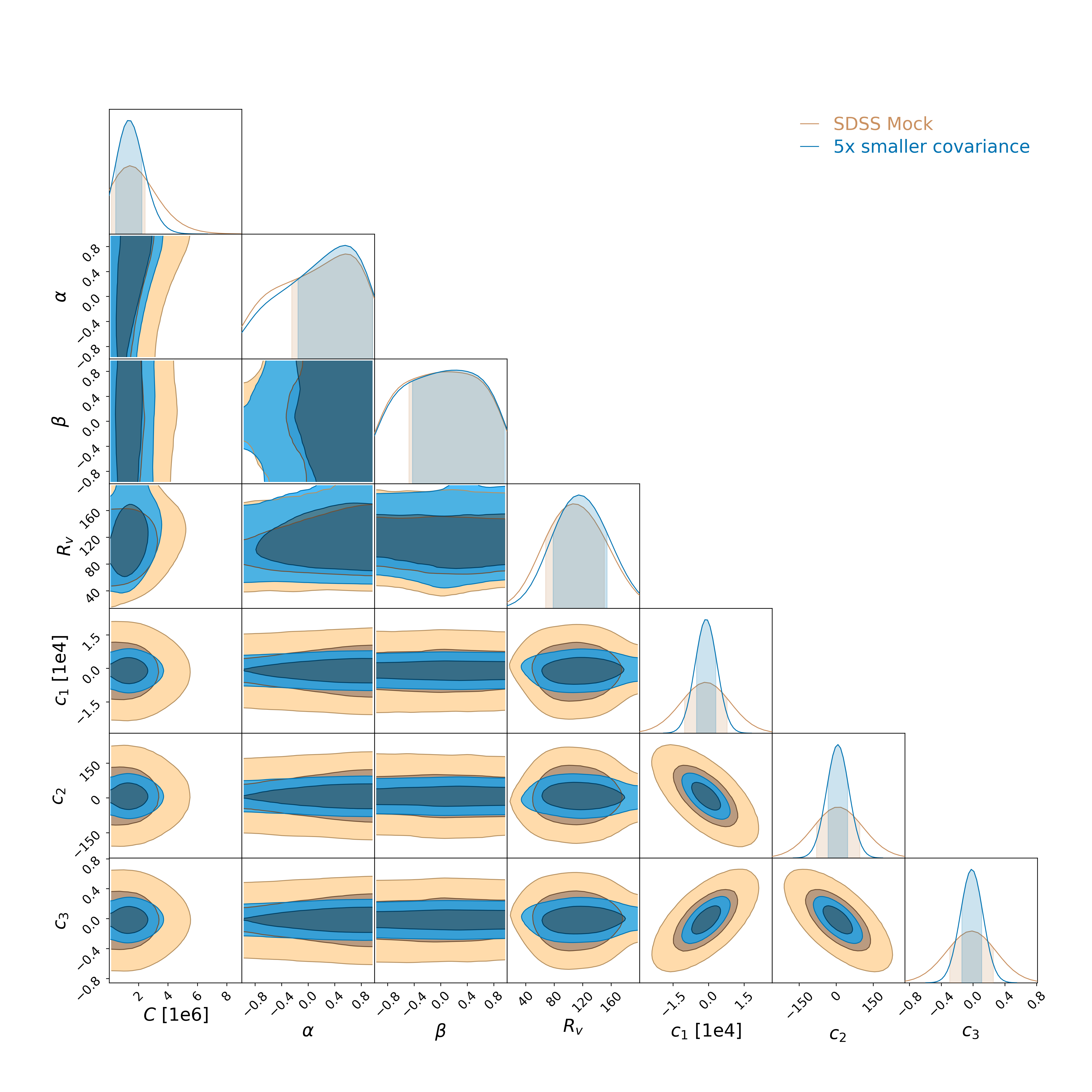}
    \caption{A comparison between the posteriors obtained using the covariance matrix of the mock realizations with and without an artificial factor of $5$ decrease in uncertainties. We notice that the constraints on the parameters in the polynomial fit of $\Psi_\parallel$ measurements increase significantly, but not much constraining power is added to the turnover fit parameters nor the {velocity coherence scale $R_v$.} }
    \label{fake_err_bar}
\end{figure}
Are such improvements feasible? As shown for example in \cite{Blake:2023kpk} (see appendix B therein for the detailed derivation), the covariance matrix of the velocity correlation function scales with the inverse volume. Therefore, PV surveys with the same precision and depth as the SDSS PV catalogue, but covering 5 times more sky area, are expected to at least mimic the constraining power of the blue contours in Fig.~\ref{fake_err_bar}. Interestingly, this is roughly the improvement in sky area one would expect combining DESI ($14000$ deg$^2$) \cite{Saulder:2023oqm, Said:2024pwm} and 4HS\footnote{\href{https://4mosthemispheresurvey.github.io/}{https://4mosthemispheresurvey.github.io/}} ($17000$ deg$^2$) \cite{2019Msngr.175....3D}. In addition to these improvements in sky coverage, these surveys will add higher number density of galaxies and richer depth (2-3 times more galaxies up to redshift $z \leq 0.15$ for DESI). This $\approx 200~ \mathrm{Mpc}/h$ increase in depth will likely allow us to extend measurements of the $\Psi$'s to higher distances, and constrain with greater precision $\beta$.\footnote{During the preparation of this paper, the DESI collaboration released in \cite{Turner:2025xpy} a measurement of the $\Psi$'s from their first data release (DR1). As discussed in the paper, these measurements extends only up to $150$ Mpc$/h$ due to the small sky coverage of the DR1 PV sample, which will however increase significantly with the DR2 data.}

We conclude that measurements of the velocity coherence scale with the methodology proposed in this paper will be feasible with greater than $20\%$ precision in the near future. An exact estimate is, however, beyond the scope of this work, {whose main objective is to provide a proof of concept of the proposed methodology}. {Future measurements} would indeed require the production of mocks for these surveys, in order to properly account for the high correlation between the  $\Psi_{\parallel}$ and $\Psi_{\perp}$ measurements. {Moreover, the robustness of the $R_v$ inference from $\Psi$'s measurements in presence of non-trivial window functions, and possible biases induced by lack of depth, different bin sizing, fitting range and alternative parametrizations to the ones in Eqs.~\eqref{turnaround},\eqref{model2} should be thouroughly analized in order to claim a robust measurement}. 

Finally, the impact of cosmic variance on peculiar velocity statistics cannot be underestimated. Since the velocity field correlation functions scale with an additional $\sim k^{-2}$ factor compared to their density contrast counterpart, one needs to probe large cosmological volumes in order to balance the cosmic variance contribution from large-scale modes ($k\ll1$). To understand the impact of cosmic variance alone on our proof of concept study, one would need to study the covariance in the velocity field statistics from a suite of mock realizations of the SDSS without any observational systematics, i.e. based on the true velocity field. Such an analysis goes beyond the scope of the present work, in which all these effects are combined together to build the covariance matrix used for the likelihood in Eq.~\eqref{likelihood}.
However, we notice that cosmic variance limitations could be overcome in the future by leveraging Kinematic Sunyaev–Zeldovich (KSZ) tomographic measurements of the velocity field at increased depth \citep{Smith:2018bpn,RiedGuachalla:2025byu}, or by combining density and velocity statistics, as suggested for example in ~\cite{Turner:2021ddy}.

\section{Discussion}\label{discussion}
This paper advocates a new choice of threshold to define the transition to statistical homogeneity based on PV measurements. As discussed in Sec.~\ref{velcohscalsec}, this can be identified with the scale $R_v$ at which the average parallel component of the peculiar velocities of galaxy pairs transitions from correlated to anti-correlated, corresponding to the zero crossing $\Psi_{\parallel}(R_v)=0$, and the change in slope of $R\Psi_\perp$. Indeed, this roughly corresponds to the scale at which the evolution of the correlation dimension $D_2$ and the scaling of the bulk in spheres $\partial_R \log{\mathcal{B}_R}$ become proportional, as shown in Fig.~\ref{RatioD/S}.

This definition has several key advantages. Unlike the scale $R_\rho$ measured from the correlation dimension $D_2$ and the mean scaled counts $\mathcal{N}_R$, it is insensitive to (and therefore not degenerate with) the galaxy bias. Furthermore, unlike $R_\rho$, the numerical value of $R_v$ is largely unaffected by the specific location and amplitude of the BAO peak, as demonstrated in Fig.~\ref{Rrhovcomparison}.

It is well known that $R_\rho$ evolves with redshift \citep{Ntelis:2017nrj,Avila:2021mbj}, as one would expect from the non-trivial redshift dependence of $\sigma_R$ and its specific numerical value. Interestingly, however, the zero crossing of $\mathcal{S}(R)$ occurs at approximately the same comoving scale at all redshifts. This can be understood by considering Eqs.~\eqref{gorskicon}, \eqref{Mateparam} and the theoretical estimator for $\Psi_\parallel$ in Eq.~\eqref{psipar}. In the linear regime, the redshift evolution contributes only an overall rescaling of the amplitude of the power spectrum $P(k)$ (in comoving Fourier scales $h/$Mpc) through the growth factor $D(z)$ and the prefactors $H^2 f^2 a^2$. This overall normalization cannot significantly change the location of the peak in $R\mathcal{B}_R$, but only its amplitude, as illustrated in Fig.~\ref{fig:hom_scale}.

These considerations show that there is no simple mapping between $R_v$ and $R_\rho$, as it is not possible to redefine the threshold used for $D_2$ so that their values match at all redshifts (which is expected, as they measure intrinsically different things). On the other hand, the proportionality between $D_2$ and the derivative of $\mathcal{B}_R$—which, unlike the thresholds $R_v$ and $R_\rho$, are the actual estimators of statistical homogeneity— is redshift independent. 

As discussed in Sec.~\ref{Theory} and Appendix~\ref{appendixB}, and explicitly shown in Fig.~\ref{rvvskeq}, $R_v$ is essentially determined by the value of $k_{eq}$. We therefore speculate that, once measured and calibrated at a given redshift, measurements of $R_v$ could be used as a standard ruler. We caution that, in order to validate this speculation, the impact of possible systematics induced by non-linearities, selection effects, and calibration biases should be thoroughly tested using large suites of cosmological simulations.

Our analysis of SDSS data shows that existing measurements can determine $R_v$ with a lower bound uncertainty of $\approx 30\%$. However, as discussed in Sec.~\ref{SDSS}, we expect that ongoing and upcoming PV surveys such as DESI, 4HS, and LSST will be able to achieve significantly higher precision in the foreseeable future.

Finally, the close analogy between Eqs.~\eqref{sigmav} and \eqref{bulkinsphere} points to a promising direction for future developments: reformulating the velocity coherence scale inference in terms of bulk-flow observables rather than $\Psi_\parallel$ and $\Psi_\perp$. Such an approach may result in alternative estimators of the velocity correlation functions, and provide a more direct link between large-scale dynamics and the emergence of cosmic homogeneity. 

\acknowledgments 
We are grateful to Sunny Vagnozzi, Stefano Camera, Madeline Lily Cross-Parkin, Eric Thrane, Valerio Marra and Agne Semenaite for useful comments and suggestions.
This research was conducted by the Australian Research Council Centre of Excellence for Gravitational Wave Discovery (project number CE230100016) and funded by the Australian Government. 

\appendix

\section{Homogeneous distributions}\label{appendixA}

{A field $\rho$ generated by a stationary stochastic process is homogeneous if its probability density functional $\mathcal{P}[\rho(\vec{x})]$ is translationally invariant. This is equivalent to demanding that its ensemble-average has a well-defined and position-independent value $\langle\rho(\vec{x})\rangle\equiv \rho_0$. Furthermore, higher order $n$ moments must depend only on the vector separations between the $n$ points. In terms of the two point correlation function, this implies
\begin{equation}
    C(\vec{x},\vec{y}) = \langle\rho(\vec{x}),\rho(\vec{y})\rangle\equiv C(\vec{x}-\vec{y})\;,
\end{equation}
which, further assuming statistical isotropy, becomes a function of the scalar separation only  $C\equiv C(r)$ (where $r=\sqrt{|\vec{x}-\vec{y}|^2}$).
Finally, a crucial assumption for cosmological applications is that the probability density $\mathcal{P}[\rho]$ is ergodic, which usually allows one to replace the ensemble average $\langle\;,\rangle$ with the spatial average over the total volume (infinite in the thermodynamic limit) of an individual realization of the field. This property is usually referred to as the self-averaging property of the distribution.
When considering an intrinsically discrete distribution of tracers, i.e. a density distribution of the form
\begin{equation}
    \rho(\vec{r}) =\sum_i^N \delta(\vec{r} - \vec{r}_i)\;,
\end{equation}
where $\delta$ is the Dirac delta function and the sum runs over the $N$ particles in the volume considered, one is usually interested in the conditional probability of observing a tracer around one centered at position $\vec{x}_o$ (such as the observer position). In this case, the correlation function can be written
\begin{equation}C(\vec{x}_0,\vec{x}_0+\vec{r}) = \frac{\delta(\vec{r)}}{\rho_0} +\xi(r)\;,
\end{equation}
where $\xi$ can be thought of as the ``off-diagonal'' part of $C(\vec{r})$, and it is physically meaningful only when $r\neq 0$.}
{
We can now define the homogeneity scale $R_H$ for a distribution which is locally inhomogeneous but that becomes homogeneous on sufficiently large scales:
\begin{equation}\label{homscaledef}
    \left| \frac{3}{R^3}\int_0^R dr \rho(r_i)r^2 - \rho_0 \right| < \rho_0\;\;\forall R\geq R_H\;, \forall r_i\;,
\end{equation}
where $r_i$ indicates the arbitrary center of a sphere anywhere within the field realization. This definition, written in terms of the correlation function $\xi(r)$, implies:
\begin{equation}
    \left|\frac{1}{R^3}\int_0^R dr\; r^2 \xi(\vec{x}_0 +\vec{r})\right| = 0 \;\forall\; R\geq R_H\;, \forall \vec{x}_0 \in \mathcal{V}\;,
\end{equation}
which is realised for example in the case of pure white noise for which $\xi(r)=0$. }

\section{Comparison of $R_v$ and $R_\rho$ across cosmologies}\label{appendixB}
{Whilst the scaling of $\mathcal{N}$ and $\mathcal{B}_R$ becomes approximately proportional asymptotically, the same proportionality does not apply to the numerical values $R_v$ and $R_\rho$. This is because whilst the numerical value of $R_v$ is determined essentially by the \textit{shape} of $P(k)$ only, $R_\rho$ is also affected by its \textit{amplitude}. This  occurs because the crossing of the constant $D_2 = 2.97$ threshold depends largely on the strength and position of the BAO peak in a given cosmology, which can either fall within or above $1\%$ tolerance. Notice that only $k$ modes relatively close to the turnover of the power spectrum impact this numerical value, so varying for example $\Omega_k$ has little impact on the value of either $R_{v,\rho}$. To illustrate this more clearly, we plot in Fig.~\ref{Rrhovcomparison} the correlation dimension for a few different cosmologies obtained by varying either one of $\Omega_m$ or$ H_0$ keeping the other fixed, or the matter radiation equality $k_{eq}\propto \omega_mh^2$ keeping the present day normalized matter density $\Omega_m$ constant.}

\begin{figure}[h]
    \centering
    % \hspace*{-1.5cm}
    \includegraphics[scale=0.6]{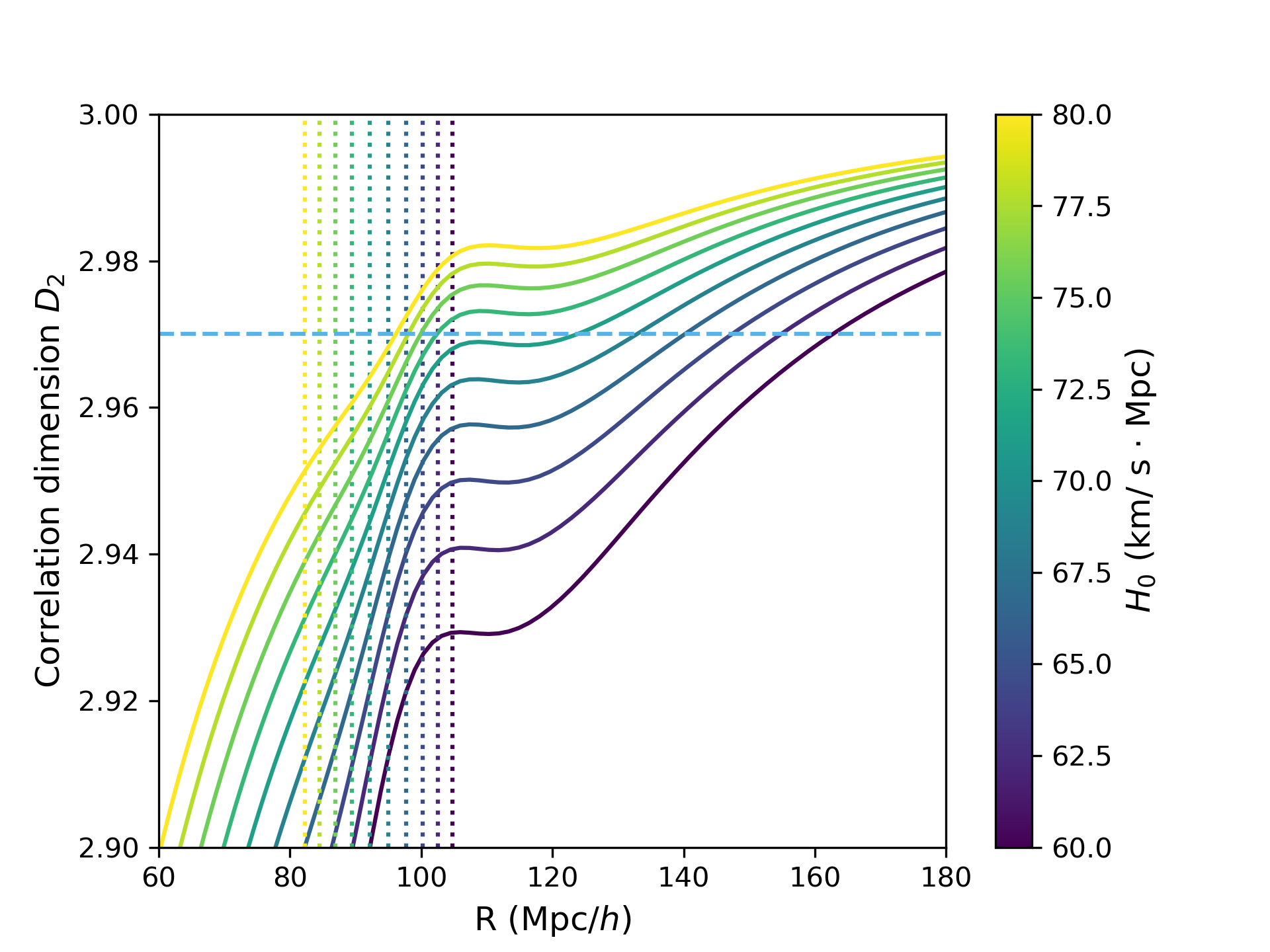}\includegraphics[scale=0.6]{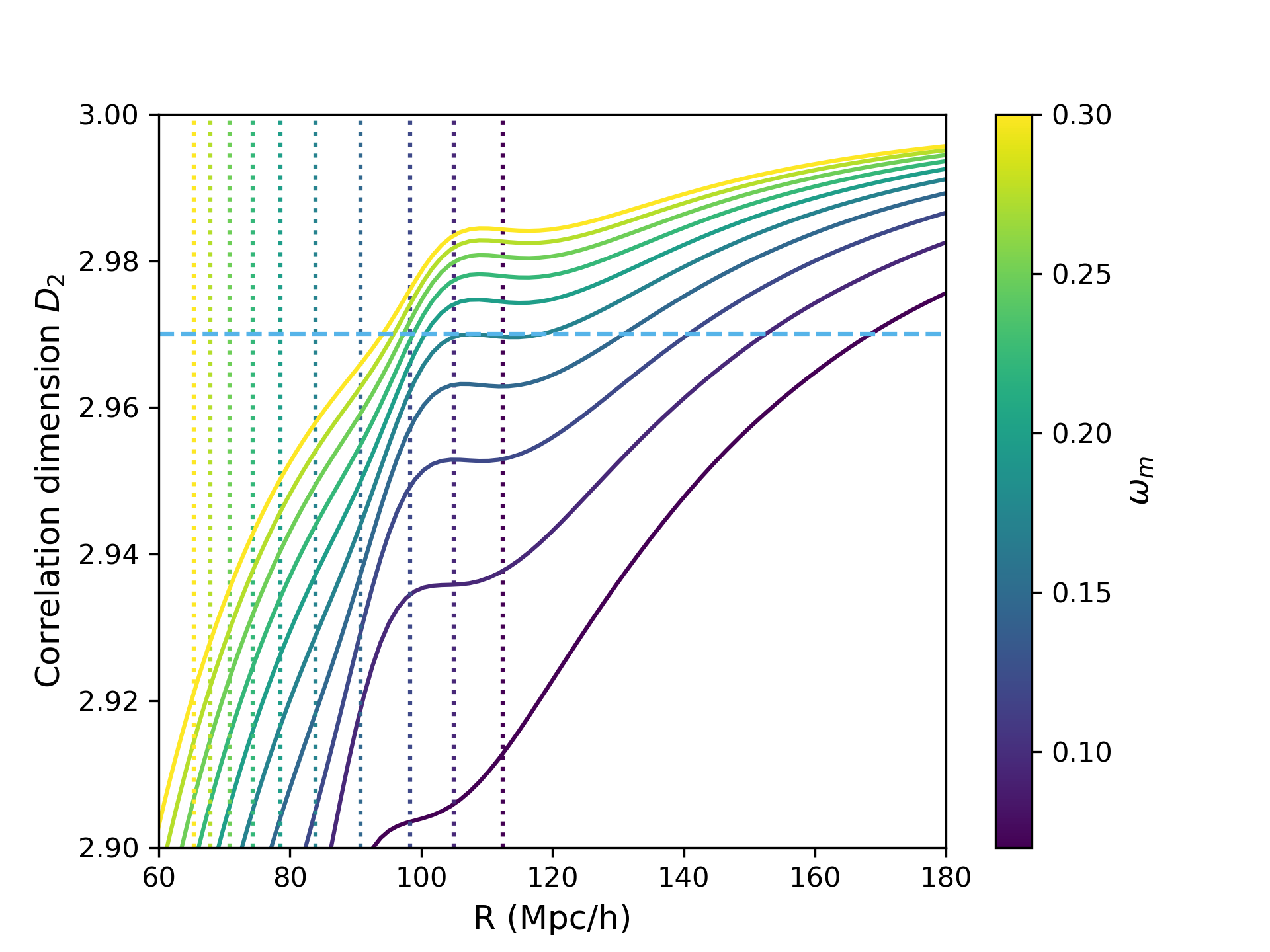}\\
    \includegraphics[scale=0.6]{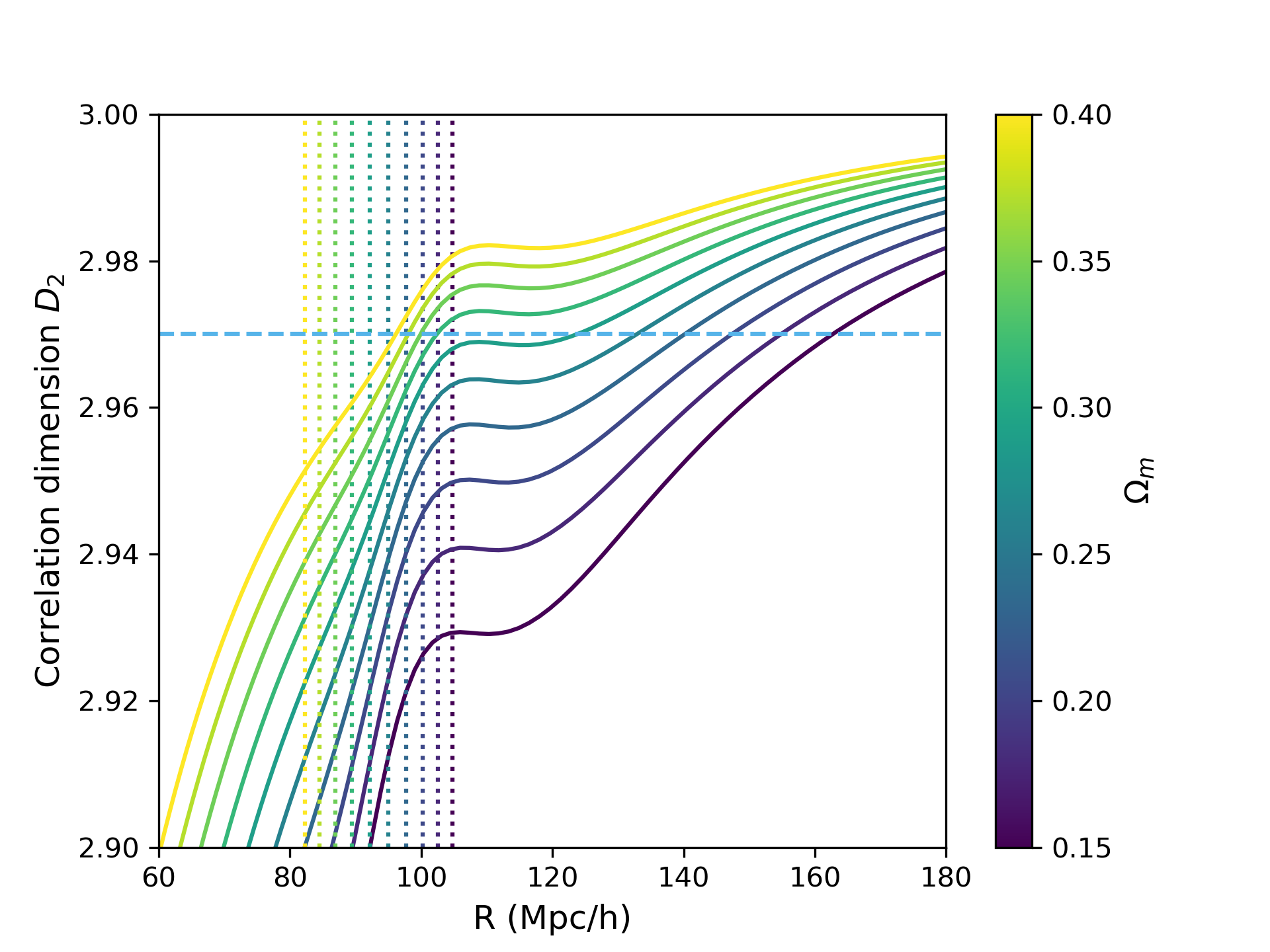}  \caption{{The correlation dimension $D_2$ for a set of fiducial $\Lambda$CDM cosmologies varying either $\Omega_m, H_0$ or $\omega_m$ (in the latter case keeping constant the normalized present day matter density $\Omega_m$). The blue dashed line shows the  $D_2=2.97$ threshold defining $R_\rho$, which, as the plot shows, is intersected by the $D_2$ curves quite differently depending on the amplitude and position of the BAO peak (despite the fact that all the distributions are statistically homogeneous, sharing the same asymptotic behaviour). On the other hand, the velocity coherence scale $R_v$ (dotted vertical lines) varies more smoothly across cosmologies.
}}
    \label{Rrhovcomparison}
\end{figure}

{We see that whilst the numerical value of $R_v$ changes smoothly with either of the parameters being varied, the threshold scale $R_\rho$ varies much more abruptly.
This is particularly important, since the conversion from redshifts to distances strongly depends on $H_0$. For example, a $ 1.5\%; $ variation $\Delta H_0$ around a fiducial value $H_0=70 \;\rm{km}/s\;\rm{Mpc}$ can cause shifts in the inferred homogeneity scale of order $\Delta R_\rho \approx 20 \rm{Mpc}$$/h$.}

{It is interesting to understand how the $R_\rho$ and $R_v$ statistics behave in truly inhomogeneous spacetimes. As an exercise, we employed the power spectra $P(k)$ computed in a suite of Lemaitre-Tolman-Bondi (LTB) simulations within the BEHOMO \footnote{\href{https://valerio-marra.github.io/BEHOMO-project//}{https://valerio-marra.github.io/BEHOMO-project//}} project, first presented in ~\cite{Marra:2022ixf}. In particular, we restrict ourselves to a set of simulations in a cubic box of 1 Gpc$^3$ volume, with a central LTB inhomogeneity of radius $r_b =400$ Mpc and different values of its average density contrast $\delta$. Fig.~ \ref{LTBsnaps} shows the snapshots at redshift $z=0$ of these simulations and the  corresponding correlation dimension $D_2$, as well as the velocity coherence scale $R_v$. We notice that the numerical values of both $R_v$ and $R_\rho$ are well below the size of the central inhomogeneity, and therefore one would wrongly conclude that the matter distribution in these boxes is homogeneous. This simple exercise, despite not being rigorous, shows that these estimators should be generally treated as a consistency test, indicating the typical scales of fluctuations in homogeneous cosmologies, rather than direct evidence of homogeneity.}
\begin{figure}[h]
    \centering
    % \hspace*{-1.5cm}
    \includegraphics[scale=0.535]{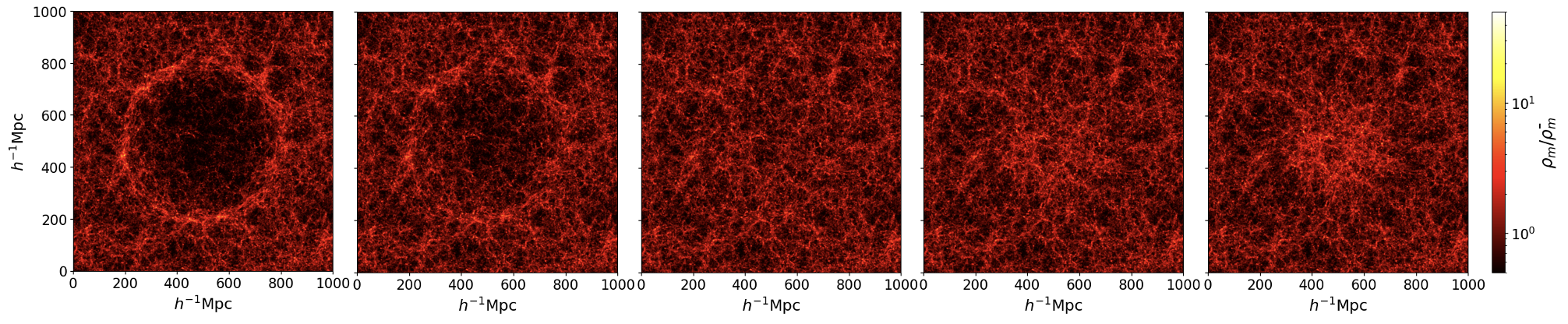}\\
    \includegraphics[scale=0.8]{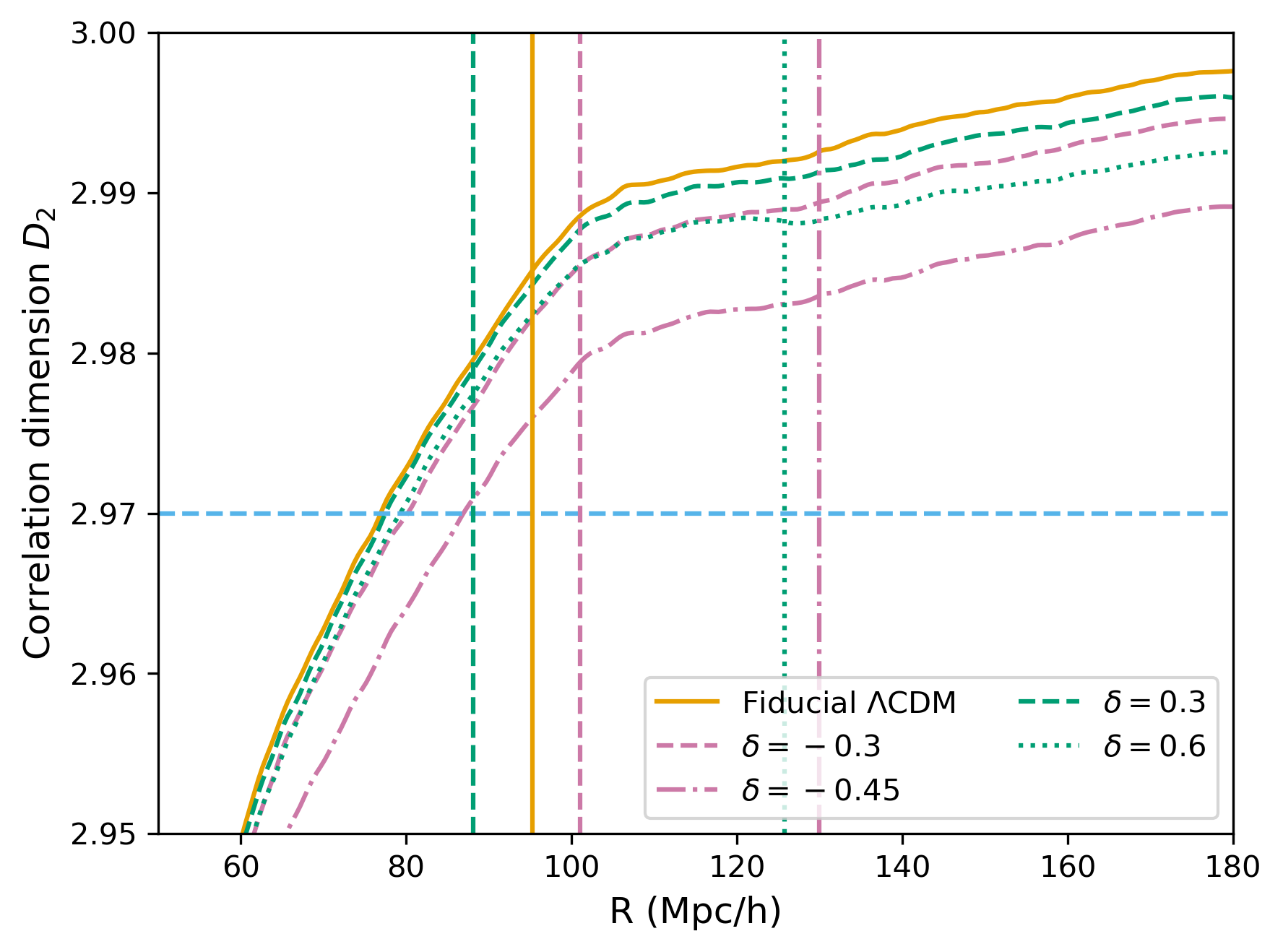}
      \caption{{\textit{Top:} Snapshots at redshift $z=0$ of the $\Lambda$LTB simulations from the BEHOMO project \citep{Marra:2022ixf} considered in this exercise. From left to right, the average density contrasts of the $\Lambda$LTB inhomogeneity are $\delta=-0.45, -0.3, 0, 0.3, 0.6$. \textit{Bottom:} The correlation dimension curves and the velocity coherence scales (vertical lines) computed from the above snapshots.
}}
    \label{LTBsnaps}
\end{figure}

\section{$R_v$ dependence on $k_{eq}$ and the impact of non-linear modes}\label{appendixC}

{A theoretical advantage of $R_v$ as a standard ruler is that it is rather simple to understand its cosmological dependence. The zero crossing of $\Psi_\parallel(R_v)=0$ is essentially determined by the shape of the power spectrum, as it is equivalent to the condition
\begin{equation}
    \int_0^\infty dk P(k)\left(j_0 - 2\frac{j_1(kR_v)}{kR_v}\right)= \int_0^\infty dk P(k)\frac{d}{d (kR_v)}\left[j_1(kR_v)\right]=0 \;.
\end{equation}
Let us now consider a triangular toy model for the power spectrum $P(k)$ such that
\begin{equation}
    P(k) \propto \begin{cases}
       k\;\;\;\;\mathrm{for} \; k< k_{eq}\\
       k^{-3} \;\mathrm{for} \; k>k_{eq}\;
    \end{cases}\;,
\end{equation}
in such a way that the integrals can be rewritten as:
\begin{equation}
    \int_0^{k_{eq}} dk\; k \frac{d}{d (kR_v)}\left[j_1(kR_v)\right] + \int_{k_{eq}}^{\infty}dk\; k^{-3}\frac{d}{d (kR_v)}\left[j_1(kR_v)\right] = 0\;.
\end{equation}
The integrals above can be solved analytically using integration by parts. A tedious calculation results in the following expression
\begin{equation}\label{rvanal}
f(R_v)=
\frac{1}{R_v^2}\left(1-\frac{\sin(y)}{y}\right)
+
\frac{3R_v^2}{30}
\left[
-\frac{(y^2+6)\cos y}{y^4}
+\operatorname{Ci}(y)
+\frac{(6+2y-y^{4})\sin y}{y^5}
\right] = 0\;,
\end{equation}
where we have defined $y=k_{\rm eq}R_v$. The functional form of the roots of the above equation is highly non-trivial, but one can nevertheless understand that the leading behaviour for $k_{eq}\rightarrow 0$ becomes $R_v\propto 1/k_{eq}$. In the opposite regime, for growing $k_{eq}$, we can see that $R_v$ behaves as some power law. However, since $k_{eq}\ll 1$, any $n>0$ polynomial of degree $n$ would be dominated by the linear term. For this reason, we propose the following fitting formula
\begin{equation}
    R_v(k_{eq}) \approx \frac{a}{k_{eq}} + bk_{eq} + c\;, 
\end{equation}
as in Eq.~\eqref{rvofkeq}. Fig.~\ref{rvvskeq} shows that this formula is relatively robust even allowing for a non-linear $P(k)$ (which CLASS computes using \texttt{halofit}). On the other hand, it is reasonable to ask whether the formula for $R_v$ we obtained from the zero crossing of $\Psi_\parallel$ still holds in the presence of non-linear modes, i.e. if Eq.~\eqref{gorskicon} is satisfied.}

{
To show that this is indeed the case, we compute in Fig.~\ref{gorskilinvsnonlin} the numerical values of $\Psi_\parallel$ and the derivative of $R\Psi_\perp$ for a fiducial non-linear cosmology. We see that whilst the differences are more pronounced in the non-linear case, they are always small and in particular below $ 10^{-2}$ across the distance range probed.}

\begin{figure}[t]
    \centering
    \hspace{1 cm}
    \includegraphics[scale=0.75]{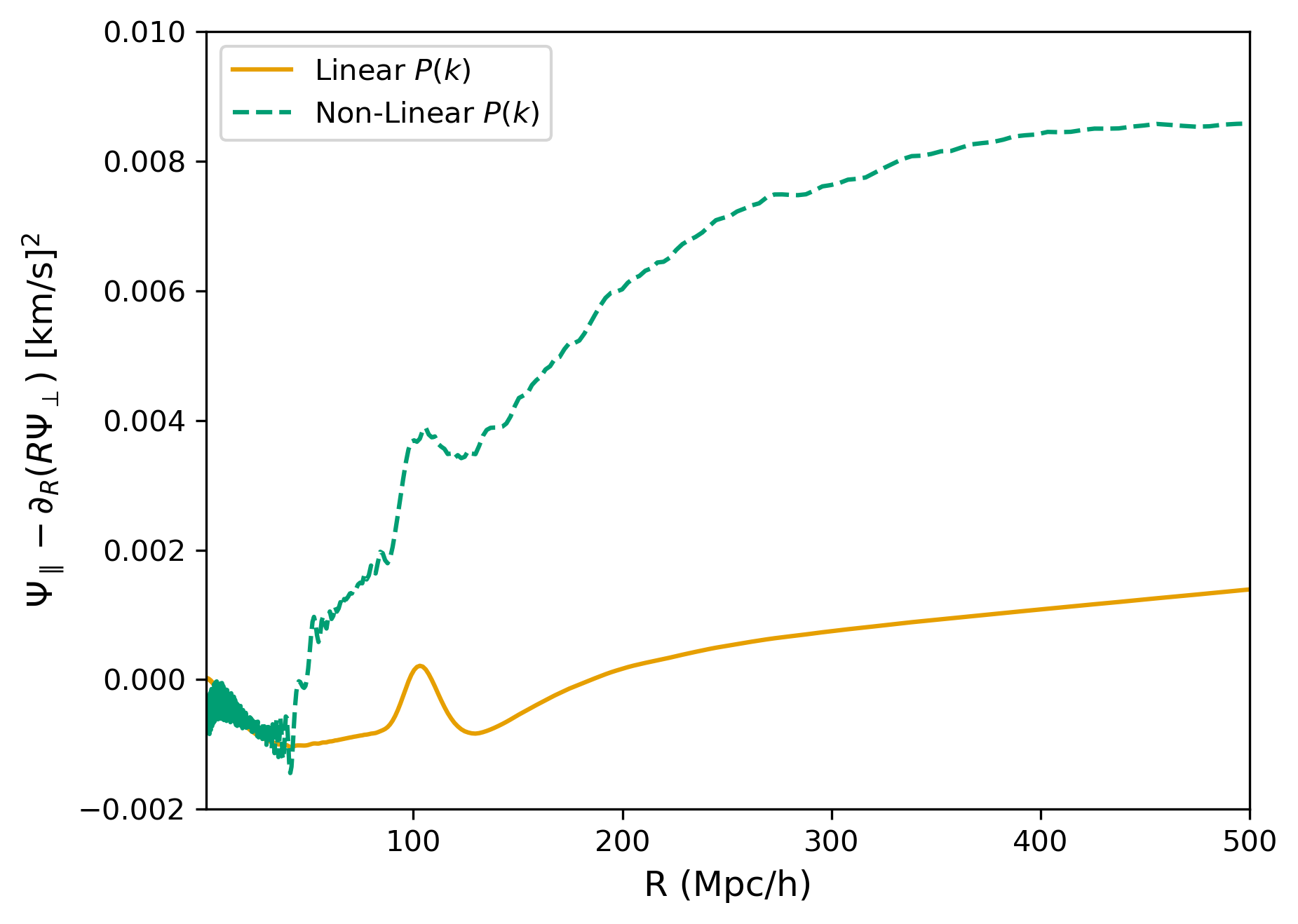}  \caption{{The numerical differences between $\Psi_\parallel$ and the derivative of $(R\Psi_\perp)$ computed using fiducial linear (orange) and non-linear (green) power spectra, showing that these are smaller than $\leq 0.01$ across the entire distance range probed.
}}
    \label{gorskilinvsnonlin}
\end{figure}
This is expected, since the only physical mechanism that can introduce differences between the two are vorticity effects, which are expected to occur only below $\leq 1 \rm{Mpc}$ scales. We conclude that the proposed methodologies to estimate $R_v$, especially given the precision of current and upcoming cosmological observations, are largely unaffected by any realistic non-linear effect.

\section{Consistency between different choices of polynomial fitting}\label{appendixD}
We verified that increasing or decreasing the order of the polynomial fitting to identify the zero-crossing in $\Psi_\parallel$ does not significantly change the marginalized posterior on $R_v$ for neither the mocks or the data considered in section~\ref{SDSS}, as clear from Fig.~\ref{fig:polynomial_degrees}.
\begin{figure}[t]
    \centering
    \hspace{1 cm}
    \includegraphics[scale=0.65]{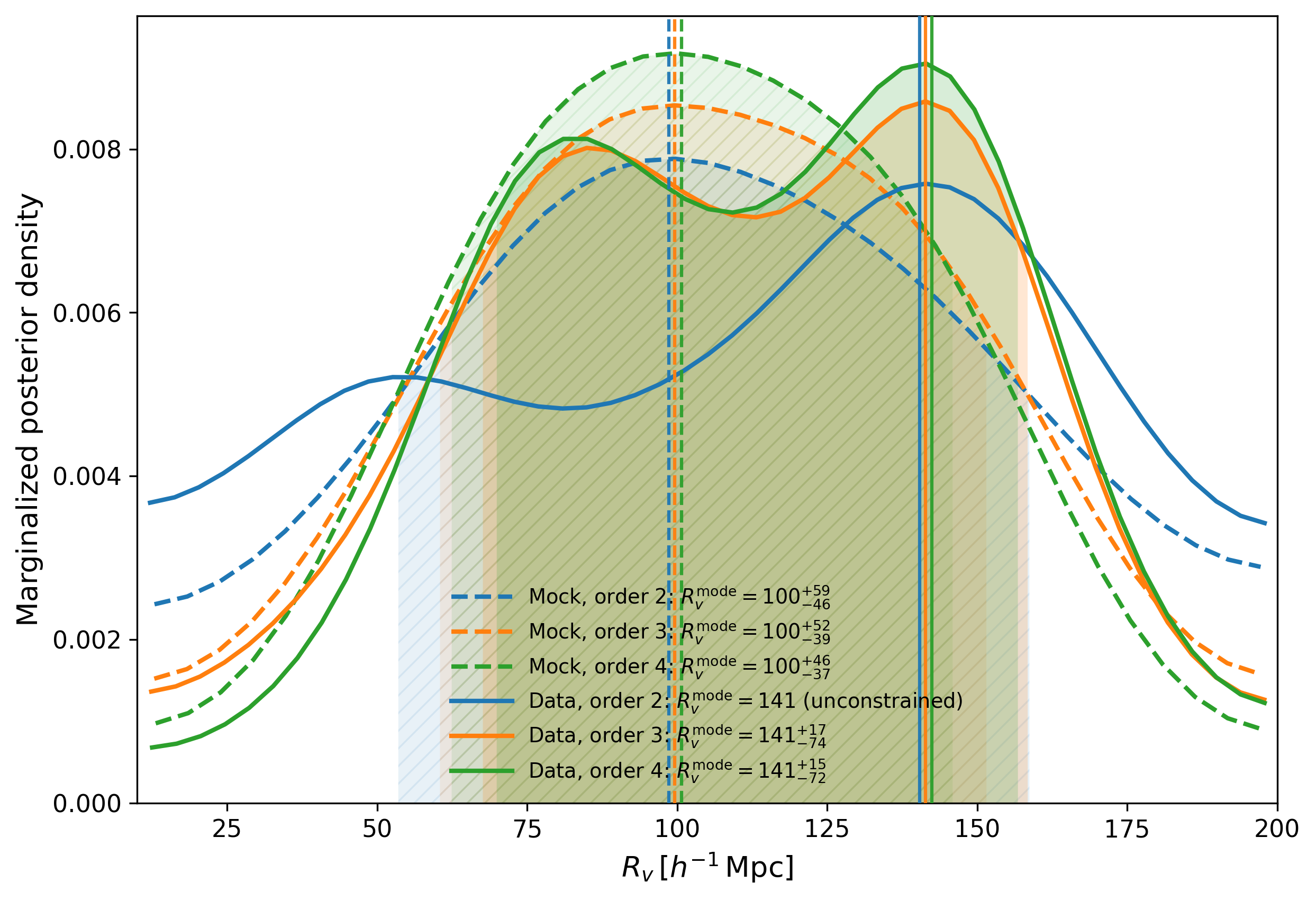}  \caption{{Marginalized posterior distributions of $R_v$ obtained by fitting $\Psi_\parallel$ with polynomials of different degrees for the SDSS data (solid lines) and the mocks (dashed lines). The broadly similar posterior distributions indicate that the inferred velocity-coherence scale is relatively insensitive to the adopted polynomial order. Vertical lines mark the marginalized modes and have been displaced by a small artificial horizontal offset for visibility. The shaded regions indicate the $68\%$ equal-density credible intervals. For the quadratic fit to the data, the broad and asymmetric posterior does not yield a closed $68\%$ equal-density interval within the adopted prior range. Consequently, only the marginalized mode is shown, and its uncertainty is reported as unconstrained.
}}
    \label{fig:polynomial_degrees}
\end{figure}

\bibliography{oja_template}

@article{Gabrielli:2000sh,
    author = "Gabrielli, A. and Sylos Labini, Francesco",
    title = "{Fluctuations in galaxy counts: a new test for homogeneity versus fractality}",
    eprint = "astro-ph/0012097",
    archivePrefix = "arXiv",
    doi = "10.1209/epl/i2001-00239-3",
    journal = "EPL",
    volume = "54",
    pages = "286",
    year = "2001"
}

@article{Gabrielli:2001xw,
    author = "Gabrielli, Andrea and Joyce, Michael and Sylos Labini, Francesco",
    title = "{The Glass - like universe: Real space correlation properties of standard cosmological models}",
    eprint = "astro-ph/0110451",
    archivePrefix = "arXiv",
    reportNumber = "LPT-ORSAY-01-92",
    doi = "10.1103/PhysRevD.65.083523",
    journal = "Phys. Rev. D",
    volume = "65",
    pages = "083523",
    year = "2002"
}

@article{Planck:2018vyg,
    author = "Aghanim, N. and others",
    collaboration = "Planck",
    title = "{Planck 2018 results. VI. Cosmological parameters}",
    eprint = "1807.06209",
    archivePrefix = "arXiv",
    primaryClass = "astro-ph.CO",
    doi = "10.1051/0004-6361/201833910",
    journal = "Astron. Astrophys.",
    volume = "641",
    pages = "A6",
    year = "2020",
    note = "[Erratum: Astron.Astrophys. 652, C4 (2021)]"
}

@article{Marra:2022ixf,
    author = "Marra, V. and Castro, T. and Camarena, D. and Borgani, S. and Ragagnin, A.",
    title = "{The BEHOMO project: \ensuremath{\Lambda} Lema\^\i{}tre-Tolman-Bondi N-body simulations}",
    eprint = "2203.04009",
    archivePrefix = "arXiv",
    primaryClass = "astro-ph.CO",
    doi = "10.1051/0004-6361/202243539",
    journal = "Astron. Astrophys.",
    volume = "664",
    pages = "A179",
    year = "2022"
}

@ARTICLE{2013PASP..125..306F,
       author = {{Foreman-Mackey}, Daniel and {Hogg}, David W. and {Lang}, Dustin and {Goodman}, Jonathan},
        title = "{emcee: The MCMC Hammer}",
      journal = {pasp},
         year = 2013,
        month = mar,
       volume = {125},
       number = {925},
        pages = {306},
          doi = {10.1086/670067},
archivePrefix = {arXiv},
       eprint = {1202.3665},
 primaryClass = {astro-ph.IM},
       adsurl = {https://ui.adsabs.harvard.edu/abs/2013PASP..125..306F}
}

@article{4HS,
  author =  "De Jong, S. and others",
  title = "{4MOST : Project overview and information for the First Call for Proposals}",
  doi = "10.18727/0722-6691/5117",
  journal = "The Messenger",
  volume = "175",
  pages = "3-11",
  year = "2019",
}

@article{Saulder:2023oqm,
    author = "Saulder, Christoph and others",
    title = "{Target Selection for the DESI Peculiar Velocity Survey}",
    eprint = "2302.13760",
    archivePrefix = "arXiv",
    primaryClass = "astro-ph.CO",
    month = "2",
    year = "2023"
}

@article{Scrimgeour:2012wt,
    author = "Scrimgeour, Morag and others",
    title = "{The WiggleZ Dark Energy Survey: the transition to large-scale cosmic homogeneity}",
    eprint = "1205.6812",
    archivePrefix = "arXiv",
    primaryClass = "astro-ph.CO",
    doi = "10.1111/j.1365-2966.2012.21402.x",
    journal = "Mon. Not. Roy. Astron. Soc.",
    volume = "425",
    pages = "116--134",
    year = "2012"
}

@article{Ntelis:2017nrj,
    author = "Ntelis, Pierros and others",
    title = "{Exploring cosmic homogeneity with the BOSS DR12 galaxy sample}",
    eprint = "1702.02159",
    archivePrefix = "arXiv",
    primaryClass = "astro-ph.CO",
    doi = "10.1088/1475-7516/2017/06/019",
    journal = "JCAP",
    volume = "06",
    pages = "019",
    year = "2017"
}

@article{Whitford:2023oww,
    author = "Whitford, A. M. and Howlett, C. and Davis, T. M.",
    title = "{Evaluating bulk flow estimators for CosmicFlows-4 measurements}",
    eprint = "2306.11269",
    archivePrefix = "arXiv",
    primaryClass = "astro-ph.CO",
    month = "6",
    year = "2023"
}

@article{Giani:2023aor,
    author = "Giani, Leonardo and Howlett, Cullan and Said, Khaled and Davis, Tamara and Vagnozzi, Sunny",
    title = "{An effective description of Laniakea: impact on cosmology and the local determination of the Hubble constant}",
    eprint = "2311.00215",
    archivePrefix = "arXiv",
    primaryClass = "astro-ph.CO",
    doi = "10.1088/1475-7516/2024/01/071",
    journal = "JCAP",
    volume = "01",
    pages = "071",
    year = "2024"
}

@ARTICLE{emcee,
       author = {{Foreman-Mackey}, Daniel and {Hogg}, David W. and {Lang}, Dustin and {Goodman}, Jonathan},
        title = "{emcee: The MCMC Hammer}",
      journal = {pasp},
         year = 2013,
        month = mar,
       volume = {125},
       number = {925},
        pages = {306},
          doi = {10.1086/670067},
archivePrefix = {arXiv},
       eprint = {1202.3665},
 primaryClass = {astro-ph.IM}
}

@article{Giani:2025hhs,
    author = "Giani, Leonardo and Von Marttens, Rodrigo and Piattella, Oliver Fabio",
    title = "{The matter with(in) CPL}",
    eprint = "2505.08467",
    archivePrefix = "arXiv",
    primaryClass = "astro-ph.CO",
    month = "5",
    year = "2025"
}

@article{Martinez:1998yp,
    author = "Martinez, Vicent J. and Pons-Borderia, Maria-Jesus and Moyeed, Rana A. and Graham, Matthew J.",
    title = "{Searching for the scale of homogeneity}",
    eprint = "astro-ph/9804073",
    archivePrefix = "arXiv",
    doi = "10.1046/j.1365-8711.1998.01730.x",
    journal = "Mon. Not. Roy. Astron. Soc.",
    volume = "298",
    pages = "1212",
    year = "1998"
}

@article{Amendola:1999gd,
    author = "Amendola, Luca and Palladino, Emilia",
    title = "{The scale of homogeneity in the las campanas redshift survey}",
    eprint = "astro-ph/9901420",
    archivePrefix = "arXiv",
    doi = "10.1086/311936",
    journal = "Astrophys. J. Lett.",
    volume = "514",
    pages = "L1",
    year = "1999"
}

@article{Pan:2000yg,
    author = "Pan, Jun and Coles, Peter",
    title = "{Large scale cosmic homogeneity from a multifractal analysis of the pscz catalogue}",
    eprint = "astro-ph/0008240",
    archivePrefix = "arXiv",
    doi = "10.1046/j.1365-8711.2000.03965.x",
    journal = "Mon. Not. Roy. Astron. Soc.",
    volume = "318",
    pages = "L51",
    year = "2000"
}

@article{Yadav:2005vv,
    author = "Yadav, Jaswant and Bharadwaj, Somnath and Pandey, Biswajit and Seshadri, T. R.",
    title = "{Testing homogeneity on large scales in the Sloan Digital Sky Survey Data Release One}",
    eprint = "astro-ph/0504315",
    archivePrefix = "arXiv",
    doi = "10.1111/j.1365-2966.2005.09578.x",
    journal = "Mon. Not. Roy. Astron. Soc.",
    volume = "364",
    pages = "601--606",
    year = "2005"
}

@ARTICLE{2009MNRAS.399L.128S,
       author = {{Sarkar}, Prakash and {Yadav}, Jaswant and {Pandey}, Biswajit and {Bharadwaj}, Somnath},
        title = "{The scale of homogeneity of the galaxy distribution in SDSS DR6}",
      journal = {mnras},
         year = 2009,
        month = oct,
       volume = {399},
       number = {1},
        pages = {L128-L131},
          doi = {10.1111/j.1745-3933.2009.00738.x},
archivePrefix = {arXiv},
       eprint = {0906.3431},
 primaryClass = {astro-ph.CO},
       adsurl = {https://ui.adsabs.harvard.edu/abs/2009MNRAS.399L.128S}
}

@ARTICLE{1994ApJ...437..550M,
       author = {{Martinez}, Vicent J. and {Coles}, Peter},
        title = "{Correlations and Scaling in the QDOT Redshift Survey}",
      journal = {\apj},
         year = 1994,
        month = dec,
       volume = {437},
        pages = {550},
          doi = {10.1086/175019},
       adsurl = {https://ui.adsabs.harvard.edu/abs/1994ApJ...437..550M}
}

@article{Labini:2010qx,
    author = "Labini, Francesco Sylos and Baryshev, Yuri V.",
    title = "{Testing the Copernican and Cosmological Principles in the local universe with galaxy surveys}",
    eprint = "1006.0801",
    archivePrefix = "arXiv",
    primaryClass = "astro-ph.CO",
    doi = "10.1088/1475-7516/2010/06/021",
    journal = "JCAP",
    volume = "06",
    pages = "021",
    year = "2010"
}

@article{Labini:2011dv,
    author = "Labini, Francesco Sylos",
    title = "{Very large scale correlations in the galaxy distribution}",
    eprint = "1110.4041",
    archivePrefix = "arXiv",
    primaryClass = "astro-ph.CO",
    doi = "10.1209/0295-5075/96/59001",
    journal = "EPL",
    volume = "96",
    number = "5",
    pages = "59001",
    year = "2011"
}

@article{Martinez:2002mi,
    author = "Martinez, Vicent J. and Saar, Enn",
    editor = "Gorhamn, Peter W.",
    title = "{Clustering statistics in cosmology}",
    eprint = "astro-ph/0209208",
    archivePrefix = "arXiv",
    doi = "10.1117/12.461972",
    journal = "Proc. SPIE Int. Soc. Opt. Eng.",
    volume = "4847",
    pages = "86",
    year = "2002"
}

@article{Ntelis:2018ctq,
    author = "Ntelis, Pierros and Ealet, Anne and Escoffier, Stephanie and Hamilton, Jean-Christophe and Hawken, Adam James and Le Goff, Jean-Marc and Rich, James and Tilquin, Andre",
    title = "{The scale of cosmic homogeneity as a standard ruler}",
    eprint = "1810.09362",
    archivePrefix = "arXiv",
    primaryClass = "astro-ph.CO",
    doi = "10.1088/1475-7516/2018/12/014",
    journal = "JCAP",
    volume = "12",
    pages = "014",
    year = "2018"
}

@article{Avila:2021mbj,
    author = "Avila, Felipe and Bernui, Armando and Nunes, Rafael C. and de Carvalho, Edilson and Novaes, Camila P.",
    title = "{The homogeneity scale and the growth rate of cosmic structures}",
    eprint = "2111.08541",
    archivePrefix = "arXiv",
    primaryClass = "astro-ph.CO",
    doi = "10.1093/mnras/stab3122",
    journal = "Mon. Not. Roy. Astron. Soc.",
    volume = "509",
    number = "2",
    pages = "2994--3003",
    year = "2021"
}

@article{Labini:2025dnc,
    author = "Labini, Francesco Sylos and Antal, Tibor",
    title = "{Large-Scale Galaxy Correlations from the DESI First Data Release}",
    eprint = "2511.21585",
    archivePrefix = "arXiv",
    primaryClass = "astro-ph.CO",
    doi = "10.1051/0004-6361/202558271",
    journal = "Astron. Astrophys.",
    volume = "707",
    pages = "A254",
    year = "2026"
}

@ARTICLE{2013MNRAS.429.1902P,
       author = {{Poole}, Gregory B. and others},
        title = "{The WiggleZ Dark Energy Survey: probing the epoch of radiation domination using large-scale structure}",
      journal = {\mnras},
         year = 2013,
        month = mar,
       volume = {429},
       number = {3},
        pages = {1902-1912},
          doi = {10.1093/mnras/sts431},
archivePrefix = {arXiv},
       eprint = {1211.5605},
 primaryClass = {astro-ph.CO},
       adsurl = {https://ui.adsabs.harvard.edu/abs/2013MNRAS.429.1902P}
}

@article{Lai:2025xxg,
    author = "Lai, Yan and Howlett, Cullan and Davis, Tamara",
    title = "{Can a multi-tracer approach improve the constraints on the turnover scale at low redshift?}",
    eprint = "2507.11823",
    archivePrefix = "arXiv",
    primaryClass = "astro-ph.CO",
    doi = "10.1088/1475-7516/2025/10/071",
    journal = "JCAP",
    volume = "10",
    pages = "071",
    year = "2025"
}

@article{Aluri:2022hzs,
    author = "Aluri, Pavan Kumar and others",
    title = "{Is the observable Universe consistent with the cosmological principle?}",
    eprint = "2207.05765",
    archivePrefix = "arXiv",
    primaryClass = "astro-ph.CO",
    doi = "10.1088/1361-6382/acbefc",
    journal = "Class. Quant. Grav.",
    volume = "40",
    number = "9",
    pages = "094001",
    year = "2023"
}

@BOOK{Peebles1980,
       author = {{Peebles}, P.~J.~E.},
        title = "{The large-scale structure of the universe}",
         year = 1980,
       adsurl = {https://ui.adsabs.harvard.edu/abs/1980lssu.book.....P}
}

@article{Clifton:2024mdy,
    author = "Clifton, Timothy and Hyatt, Neil",
    title = "{A Radical Solution to the Hubble Tension Problem}",
    eprint = "2404.08586",
    archivePrefix = "arXiv",
    primaryClass = "astro-ph.CO",
    month = "4",
    year = "2024",
    journal = "preprint"
}

@ARTICLE{2011arXiv1104.2932L,
       author = {{Lesgourgues}, Julien},
        title = "{The Cosmic Linear Anisotropy Solving System (CLASS) I: Overview}",
      journal = {arXiv e-prints},
         year = 2011,
        month = apr,
          eid = {arXiv:1104.2932},
        pages = {arXiv:1104.2932},
          doi = {10.48550/arXiv.1104.2932},
archivePrefix = {arXiv},
       eprint = {1104.2932},
 primaryClass = {astro-ph.IM},
       adsurl = {https://ui.adsabs.harvard.edu/abs/2011arXiv1104.2932L}
}

@article{Qin:2019axr,
    author = "Qin, Fei and Howlett, Cullan and Staveley-Smith, Lister",
    title = "{The redshift-space momentum power spectrum \textendash{} II. Measuring the growth rate from the combined 2MTF and 6dFGSv surveys}",
    eprint = "1906.02874",
    archivePrefix = "arXiv",
    primaryClass = "astro-ph.CO",
    doi = "10.1093/mnras/stz1576",
    journal = "Mon. Not. Roy. Astron. Soc.",
    volume = "487",
    number = "4",
    pages = "5235--5247",
    year = "2019"
}

@article{Giani:2024nnv,
    author = "Giani, Leonardo and Von Marttens, Rodrigo and Camilleri, Ryan",
    title = "{Novel Approach to Cosmological Nonlinearities as an Effective Fluid}",
    eprint = "2410.15295",
    archivePrefix = "arXiv",
    primaryClass = "astro-ph.CO",
    doi = "10.1103/zr92-m7py",
    journal = "Phys. Rev. Lett.",
    volume = "135",
    number = "7",
    pages = "071004",
    year = "2025"
}

@article{Shao:2025xgi,
    author = "Shao, Xiaoyun and Gon{\c{c}}alves, Rodrigo and Bengaly, Carlos A. P. and Carvalho, Gabriela C. and Alcaniz, Jailson",
    title = "{Cosmic homogeneity: the effect of redshift-space distortions and bias and cosmological constraints}",
    eprint = "2507.18720",
    archivePrefix = "arXiv",
    primaryClass = "astro-ph.CO",
    month = "7",
    year = "2025"
}

@article{Borgani:1994uy,
    author = "Borgani, Stefano",
    title = "{Scaling in the universe}",
    eprint = "astro-ph/9404054",
    archivePrefix = "arXiv",
    doi = "10.1016/0370-1573(94)00073-C",
    journal = "Phys. Rept.",
    volume = "251",
    pages = "1--152",
    year = "1995"
}

@ARTICLE{gorski,
       author = {{Gorski}, Krzysztof},
        title = "{On the Pattern of Perturbations of the Hubble Flow}",
      journal = {apjl},
         year = 1988,
        month = sep,
       volume = {332},
        pages = {L7},
          doi = {10.1086/185255},
       adsurl = {https://ui.adsabs.harvard.edu/abs/1988ApJ...332L...7G}
}

@ARTICLE{Groth1989,
       author = {{Groth}, Edward J. and {Juszkiewicz}, Roman and {Ostriker}, Jeremiah P.},
        title = "{An Estimate of the Velocity Correlation Tensor: Cosmological Implications}",
      journal = {\apj},
         year = 1989,
        month = nov,
       volume = {346},
        pages = {558},
          doi = {10.1086/168038},
       adsurl = {https://ui.adsabs.harvard.edu/abs/1989ApJ...346..558G}
}

@ARTICLE{Wang2018,
       author = {{Wang}, Yuyu and {Rooney}, Christopher and {Feldman}, Hume A. and {Watkins}, Richard},
        title = "{The peculiar velocity correlation function}",
      journal = {\mnras},
         year = 2018,
        month = nov,
       volume = {480},
       number = {4},
        pages = {5332-5341},
          doi = {10.1093/mnras/sty2224},
archivePrefix = {arXiv},
       eprint = {1808.07543},
 primaryClass = {astro-ph.CO},
       adsurl = {https://ui.adsabs.harvard.edu/abs/2018MNRAS.480.5332W}
}

@ARTICLE{Wang2021,
       author = {{Wang}, Yuyu and {Peery}, Sarah and {Feldman}, Hume A. and {Watkins}, Richard},
        title = "{Improved Methods for Estimating Peculiar Velocity Correlation Functions Using Volume Weighting}",
      journal = {\apj},
         year = 2021,
        month = sep,
       volume = {918},
       number = {2},
          eid = {49},
        pages = {49},
          doi = {10.3847/1538-4357/ac0e37},
archivePrefix = {arXiv},
       eprint = {2108.08036},
 primaryClass = {astro-ph.CO},
       adsurl = {https://ui.adsabs.harvard.edu/abs/2021ApJ...918...49W}
}

@article{Turner:2024blz,
    author = "Turner, Ryan J.",
    title = "{Cosmology with Peculiar Velocity Surveys}",
    eprint = "2411.19484",
    archivePrefix = "arXiv",
    primaryClass = "astro-ph.CO",
    month = "11",
    year = "2024"
}

@article{Turner:2022mla,
    author = "Turner, Ryan J. and Blake, Chris and Ruggeri, Rossana",
    title = "{A local measurement of the growth rate from peculiar velocities and galaxy clustering correlations in the 6dF Galaxy Survey}",
    eprint = "2207.03707",
    archivePrefix = "arXiv",
    primaryClass = "astro-ph.CO",
    doi = "10.1093/mnras/stac3256",
    journal = "Mon. Not. Roy. Astron. Soc.",
    volume = "518",
    number = "2",
    pages = "2436--2452",
    year = "2022"
}

@article{Lyall:2024mks,
    author = "Lyall, Stuart and Blake, Chris and Turner, Ryan J.",
    title = "{Constraining modified gravity scenarios with the 6dFGS and SDSS galaxy peculiar velocity data sets}",
    eprint = "2407.18684",
    archivePrefix = "arXiv",
    primaryClass = "astro-ph.CO",
    doi = "10.1093/mnras/stae1718",
    journal = "Mon. Not. Roy. Astron. Soc.",
    volume = "532",
    number = "4",
    pages = "3972--3984",
    year = "2024"
}

@article{Hinton2016, doi = {10.21105/joss.00045}, url = {https://doi.org/10.21105/joss.00045}, year = {2016}, publisher = {The Open Journal}, volume = {1}, number = {4}, pages = {45}, author = {Hinton, Samuel}, title = {ChainConsumer}, journal = {Journal of Open Source Software} }

@article{Blake:2004tr,
    author = "Blake, Chris and Bridle, Sarah",
    title = "{Cosmology with photometric redshift surveys}",
    eprint = "astro-ph/0411713",
    archivePrefix = "arXiv",
    doi = "10.1111/j.1365-2966.2005.09526.x",
    journal = "Mon. Not. Roy. Astron. Soc.",
    volume = "363",
    pages = "1329--1348",
    year = "2005"
}

@article{Howlett:2022len,
    author = "Howlett, Cullan and Said, Khaled and Lucey, John R. and Colless, Matthew and Qin, Fei and Lai, Yan and Tully, R. Brent and Davis, Tamara M.",
    title = "{The sloan digital sky survey peculiar velocity catalogue}",
    eprint = "2201.03112",
    archivePrefix = "arXiv",
    primaryClass = "astro-ph.CO",
    doi = "10.1093/mnras/stac1681",
    journal = "Mon. Not. Roy. Astron. Soc.",
    volume = "515",
    number = "1",
    pages = "953--976",
    year = "2022"
}

@article{Blake:2023kpk,
    author = "Blake, Chris and Turner, Ryan J.",
    title = "{On the correlations of galaxy peculiar velocities and their covariance}",
    eprint = "2308.15735",
    archivePrefix = "arXiv",
    primaryClass = "astro-ph.CO",
    doi = "10.1093/mnras/stad3217",
    journal = "Mon. Not. Roy. Astron. Soc.",
    volume = "527",
    number = "1",
    pages = "501--520",
    year = "2023"
}

@article{Said:2024pwm,
    author = "Said, Khaled and others",
    title = "{DESI Peculiar Velocity Survey -- Fundamental Plane}",
    eprint = "2408.13842",
    archivePrefix = "arXiv",
    primaryClass = "astro-ph.CO",
    doi = "10.1093/mnras/staf700",
    journal = "Mon. Not. Roy. Astron. Soc.",
    volume = "539",
    number = "4",
    pages = "3627--3644",
    year = "2025"
}

@ARTICLE{2019Msngr.175....3D,
       author = {{de Jong}, et al. },
        title = "{4MOST: Project overview and information for the First Call for Proposals}",
      journal = {The Messenger},
         year = 2019,
        month = mar,
       volume = {175},
        pages = {3-11},
          doi = {10.18727/0722-6691/5117},
archivePrefix = {arXiv},
       eprint = {1903.02464},
 primaryClass = {astro-ph.IM},
       adsurl = {https://ui.adsabs.harvard.edu/abs/2019Msngr.175....3D}
}

@ARTICLE{2009A&A...505..981S,
       author = {{Sylos Labini}, F. and {Vasilyev}, N.~L. and {Baryshev}, Y.~V. and {L{\'o}pez-Corredoira}, M.},
        title = "{Absence of anti-correlations and of baryon acoustic oscillations in the galaxy correlation function from the Sloan Digital Sky Survey data release 7}",
      journal = {\aap},
         year = 2009,
        month = oct,
       volume = {505},
       number = {3},
        pages = {981-990},
          doi = {10.1051/0004-6361/200911987},
archivePrefix = {arXiv},
       eprint = {0903.0950},
 primaryClass = {astro-ph.CO},
       adsurl = {https://ui.adsabs.harvard.edu/abs/2009A&A...505..981S}
}

@ARTICLE{2009A&A...496....7S,
       author = {{Sylos Labini}, F. and {Vasilyev}, N.~L. and {Baryshev}, Y.~V.},
        title = "{Large-scale fluctuations in the distribution of galaxies from the two-degree galaxy redshift survey}",
      journal = {\aap},
         year = 2009,
        month = mar,
       volume = {496},
       number = {1},
        pages = {7-23},
          doi = {10.1051/0004-6361:200810575},
archivePrefix = {arXiv},
       eprint = {0902.0229},
 primaryClass = {astro-ph.CO},
       adsurl = {https://ui.adsabs.harvard.edu/abs/2009A&A...496....7S}
}

@article{Andersen:2016ywu,
    author = "Andersen, Per and Davis, Tamara M. and Howlett, Cullan",
    title = "{Cosmology with Peculiar Velocities: Observational Effects}",
    eprint = "1609.04022",
    archivePrefix = "arXiv",
    primaryClass = "astro-ph.CO",
    doi = "10.1093/mnras/stw2252",
    journal = "Mon. Not. Roy. Astron. Soc.",
    volume = "463",
    number = "4",
    pages = "4083--4092",
    year = "2016"
}

@ARTICLE{2012ApJ...761..151L,
       author = {{Li}, Ming and {Pan}, Jun and {Gao}, Liang and {Jing}, Yipeng and {Yang}, Xiaohu and {Chi}, Xuebin and {Feng}, Longlong and {Kang}, Xi and {Lin}, Weipeng and {Shan}, Guihua and {Wang}, Long and {Zhao}, Donghai and {Zhang}, Pengjie},
        title = "{Bulk Flow of Halos in {\ensuremath{\Lambda}}CDM Simulation}",
      journal = {\apj},
         year = 2012,
        month = dec,
       volume = {761},
       number = {2},
          eid = {151},
        pages = {151},
          doi = {10.1088/0004-637X/761/2/151},
archivePrefix = {arXiv},
       eprint = {1207.5338},
 primaryClass = {astro-ph.CO},
       adsurl = {https://ui.adsabs.harvard.edu/abs/2012ApJ...761..151L}
}

@article{Watkins_2023,
   title={Analysing the large-scale bulk flow using cosmicflows4: increasing tension with the standard cosmological model},
   volume={524},
   doi={10.1093/mnras/stad1984},
   number={2},
   journal={\mnras},
   author={Watkins, Richard and others},
   year={2023},
   pages={1885–1892} 
}

@article{Lane:2023ndt,
    author = "Lane, Zachary G. and Seifert, Antonia and Ridden-Harper, Ryan and Wiltshire, David L.",
    title = "{Cosmological foundations revisited with Pantheon+}",
    eprint = "2311.01438",
    archivePrefix = "arXiv",
    primaryClass = "astro-ph.CO",
    doi = "10.1093/mnras/stae2437",
    journal = "Mon. Not. Roy. Astron. Soc.",
    volume = "536",
    number = "2",
    pages = "1752--1777",
    year = "2025"
}

@article{Turner:2021ddy,
    author = "Turner, Ryan J. and Blake, Chris and Ruggeri, Rossana",
    title = "{Improving estimates of the growth rate using galaxy{\textendash}velocity correlations: a simulation study}",
    eprint = "2101.09026",
    archivePrefix = "arXiv",
    primaryClass = "astro-ph.CO",
    doi = "10.1093/mnras/stab212",
    journal = "Mon. Not. Roy. Astron. Soc.",
    volume = "502",
    number = "2",
    pages = "2087--2096",
    year = "2021"
}

@article{Turner:2025xpy,
    author = "Turner, R. J. and others",
    title = "{The DESI DR1 Peculiar Velocity Survey: growth rate measurements from galaxy and momentum correlation functions}",
    eprint = "2512.03230",
    archivePrefix = "arXiv",
    primaryClass = "astro-ph.CO",
    month = "12",
    year = "2025"
}

@article{Watkins:2025mlc,
    author = "Watkins, Richard and Feldman, Hume A.",
    title = "{The Origins of the Bulk flow}",
    eprint = "2512.03168",
    archivePrefix = "arXiv",
    primaryClass = "astro-ph.CO",
    month = "12",
    year = "2025"
}

@article{Bizarria:2025jdh,
    author = "Bizarria, Bruno B. and Novaes, Camila P. and Avila, Felipe and Mokeddem, Rahima and da Costa, Helissa H. and Wuensche, Carlos A. and Silva, Gabriel A. S.",
    title = "{Accessing the homogeneity scale with 21 cm intensity mapping surveys}",
    eprint = "2511.13931",
    archivePrefix = "arXiv",
    primaryClass = "astro-ph.CO",
    month = "11",
    year = "2025"
}

@article{Galoppo:2025hzy,
    author = "Galoppo, Marco and Giani, Leonardo and Hills, Morag and Valade, Aur{\'e}lien",
    title = "{An effective $\Lambda$-Szekeres modelling of the local Universe with Cosmicflows-4}",
    eprint = "2512.16591",
    archivePrefix = "arXiv",
    primaryClass = "astro-ph.CO",
    month = "12",
    year = "2025"
}

@article{Camarena:2025upt,
    author = "Camarena, David and Greene, Kylar and Houghteling, John and Cyr-Racine, Francis-Yan",
    title = "{Designing concordant distances in the age of precision cosmology: The impact of density fluctuations}",
    eprint = "2507.17969",
    archivePrefix = "arXiv",
    primaryClass = "astro-ph.CO",
    doi = "10.1103/1gj5-3x4m",
    journal = "Phys. Rev. D",
    volume = "112",
    number = "8",
    pages = "083526",
    year = "2025"
}

@article{Smith:2018bpn,
    author = {Smith, Kendrick M. and Madhavacheril, Mathew S. and M{\"u}nchmeyer, Moritz and Ferraro, Simone and Giri, Utkarsh and Johnson, Matthew C.},
    title = "{Kinetic Sunyaev-Zeldovich tomography and the bispectrum}",
    eprint = "1810.13423",
    archivePrefix = "arXiv",
    primaryClass = "astro-ph.CO",
    doi = "10.1103/xvc3-rd9r",
    journal = "Phys. Rev. D",
    volume = "113",
    number = "8",
    pages = "083532",
    year = "2026"
}

@article{RiedGuachalla:2025byu,
    author = "Ried Guachalla, Bernardita and others",
    title = "{Backlighting extended gas halos around luminous red galaxies: Kinematic Sunyaev-Zel{\textquoteright}dovich effect from DESI Y1 and ACT data}",
    eprint = "2503.19870",
    archivePrefix = "arXiv",
    primaryClass = "astro-ph.GA",
    reportNumber = "FERMILAB-PUB-25-0196-PPD",
    doi = "10.1103/lqbj-wcqj",
    journal = "Phys. Rev. D",
    volume = "112",
    number = "10",
    pages = "103512",
    year = "2025"
}

\end{document}